\documentclass[epj]{webofc}
\usepackage[varg]{txfonts}   
\usepackage{amssymb,amsmath,epsfig,bm,pifont}
\usepackage[utf8]{inputenc}
\wocname{EPJ Web of Conferences}
\woctitle{CONF12}
\newcommand{\vecb}[1]{\mbox{\boldmath $#1$}}

\newcommand{\sandwich}[3]{\left< #1 \right | #2 \left | #3 \right >}

\def \le { \left    }
\def \ri { \right }

\def\({\left(}
\def\[{\left[}
\def\){\right)}
\def\]{\right]}

\newcommand{\nn}{\nonumber}

\newcommand{\bn}{{\bar n}}

\newcommand{\cd}{\cdot}

\newcommand{\ot}{\leftarrow}

\begin{document}
\selectlanguage{english}
\title{Nucleon tomography. What can we do better today than Rutherford
       100 years ago?}
%
%

\author{N.~G.~Stefanis\inst{1}\fnsep\thanks{\email{stefanis@tp2.ruhr-uni-bochum.de}} (chair) \and
        Constantia Alexandrou\inst{2,3} \fnsep\thanks{\email{alexand@ucy.ac.cy}} \and
        Tanja Horn\inst{4,5} \fnsep\thanks{\email{hornt@cua.edu}} \and
        Herv\'{e} Moutarde\inst{6} \fnsep\thanks{\email{herve.moutarde@cea.fr}} \and
        Ignazio Scimemi\inst{7} \fnsep\thanks{\email{ignazios@fis.ucm.es}}
}

\institute{Institut f\"{u}r Theoretische Physik II,
           Ruhr-Universit\"{a}t Bochum,
           D-44780 Bochum, Germany
\and
           Department of Physics, University of Cyprus, P.O. Box 20537, 1678 Nicosia, Cyprus
\and
           Computation-based Science and Technology Research Center,
           The Cyprus Institute, 20 Kavafi Str., Nicosia 2121, Cyprus
\and
           The Catholic University of America, Washington, DC 20064, USA
\and
           Jefferson Laboratory, Newport News, VA 23606, USA
\and
           IRFU, CEA, Universit\'e Paris-Saclay, F-91191 Gif-sur-Yvette, France
\and
           Departamento de F\'{i}sica Te\'{o}rica II, Universidad Complutense de Madrid,
           Ciudad Universitaria, 28040 Madrid, Spain
}

\abstract{%
A survey is presented on the current status of 3D nucleon tomography.
Several research frontiers are addressed that dominate modern physics
from theory to current and future experiments.
We have now a much more detailed spatial image of the nucleon thanks
to various theoretical concepts and methods to describe its charge
distribution and spin decomposition which are highlighted here.
The progress of lattice computations of these quantities is reported
and the prospects of what we can come to expect in the near future are
discussed.
Multi-dimensional maps of the nucleon's partonic structure appear now
within reach of forthcoming experiments.
}
\maketitle

\section{Introductory remarks}
\label{sec:intro}
The strive to understand matter in terms of elementary constituents
(``atoms'')
and bring order into the natural world lasts over thousands of years.
While the ancient Greeks invented philosophical ideas about the atoms,
their physical discovery became possible only after the invention of the
Rutherford atomic model in 1911.
This model is based on the assumption that the atom consists of a central
large mass with positive charge surrounded by rotating low-mass electrons.
The next decisive step was the observation around 1920 that the Hydrogen
nucleus can be regarded as the fundamental block of all heavier nuclei,
thus giving rise to the notion of proton.
In order to compensate for the repulsive effects of the positive charges of
the protons, Rutherford postulated the existence of neutrons which have no
electric charge but contribute to the nuclear force.
They were discovered later in experiment by his associate James Chadwick.

But even the proton and the neutron are composite particles and have an
internal structure themselves, as it was shown by Hofstadter in 1950 by
means of high-energy electron-scattering from nuclei
\cite{Hofstadter:1956qs}.
The differential cross section of the hard process
$ep \longrightarrow e'p'$ --- illustrated in Fig.\ \ref{fig:scattering}
(left) --- reveals that the proton is not a point-like object but bears an
internal structure which can be described in terms of the charge and
current form factors
$F_1(t=Q^2)$ and $F_2(t=Q^2)$,
respectively, where $Q^2$ is the momentum transfer
(i.e., the virtuality of the exchanged highly off-shell photon $-q^2=Q^2$)
in the spacelike region).
These form factors are defined in terms of the hadronic matrix element of the electromagnetic
current $V_\mu$ in the Dirac parametrization
\begin{eqnarray}
  \langle N(p',s')|V_{\mu}(x)|N(p,s)\rangle
=
  \bar{u}(p',s')\left[
                 \gamma_\mu F_1(Q^2) - \sigma_{\mu\nu}\frac{q_\nu}{2m_\text{N}} F_2(Q^2)
           \right]
  u(p,s) \, ,
\label{eq:Dirac-FF}
\end{eqnarray}
where $p$, $s$, $p'$, and $s'$ are, respectively, the momenta and spins of the incoming
and outgoing nucleons, $u's$ their spinors, and $m_\text{N}$ is the nucleon mass.
These form factors are related to the magnetic ($G_\text{M}$) and electric ($G_\text{E}$)
Sachs form factors entering the electron-proton scattering cross section by means of
the Rosenbluth formula:
\begin{eqnarray}
  G_\text{M}(Q^2)
& \!\!\! = \!\!\! &
  F_1(Q^2) + F_2(Q^2)
\\
  G_\text{E}
& \!\!\! = \!\!\! &
  F_1(Q^2) + \frac{Q^2}{\left(2m_\text{N}\right)^2}F_2(Q^2) \, .
\label{eq:Sachs-FF}
\end{eqnarray}
The above form factors can be measured in experiments
(see Sec.\ \ref{sec:horn})
and are also calculable on the lattice (see Sec.\ \ref{sec:alexandrou}).
More detailed theoretical analysis of the nucleon form factors computed
in terms of nonperturbative nucleon distribution amplitudes in convolution
with hard partonic subprocesses amenable to QCD perturbation theory can be
found in \cite{Lepage:1980fj,Chernyak:1983ej,Stefanis:1999wy}.
An extensive review which covers the comparison with the most
recent experimental data is given in \cite{Brambilla:2014jmp}.
In this report, the focus is on more recent theoretical formulations,
which go beyond the longitudinal description of the nucleon, and their
verification by measurements at current and planned experiments.

\section{Benchmarks of nucleon tomography}
\label{sec:theor-tools}
This section addresses the main theoretical framework to
describe the internal structure of the nucleon within QCD.
More detailed accounts are given in the subsequent sections.

\begin{figure}[h]                                                      
\centering
\includegraphics[width=8cm,clip]{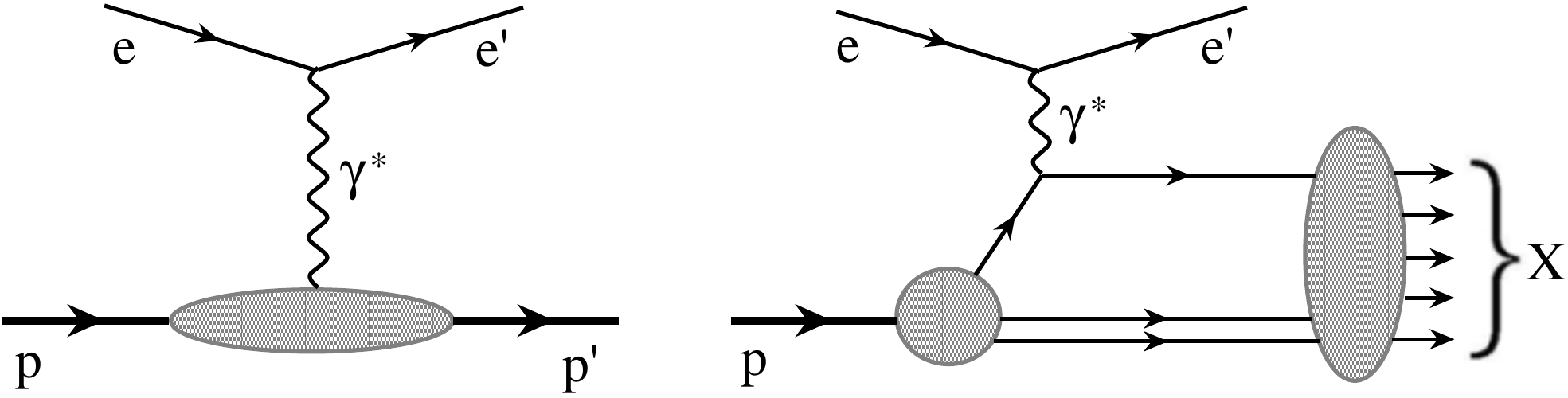} 
\caption{Left panel: illustration of elastic-electron-nucleon scattering
by means of the interaction with the virtual photon $\gamma^*(Q^2)$.
Right panel: inelastic electron-proton scattering $ep \longrightarrow e'X$
in naive parton model approximation, where the virtual photon couples to a
single parton (quark).
The produced final-state hadrons are denoted by $X$.}
\label{fig:scattering}
\end{figure}

The nucleon form factors in either representation --- Dirac or Sachs ---
parameterize in some sense our ignorance about the internal binding
effects of the nucleon which give rise to a ``diffuse'' structure
(represented by a shaded oval in the left panel of Fig.\ \ref{fig:scattering})
and cause the elastic scattering cross section to decrease with
increasing $Q^2$, hence indicating that the nucleon cannot be pointlike
\cite{Coward:1967au}.
Indeed, highly inelastic electron-proton scattering
\cite{Bloom:1969kc,Breidenbach:1969kd}
has revealed that the proton contains pointlike constituents ---
partons --- which couple to the probing highly virtual photon
(right panel in Fig.\ \ref{fig:scattering}).

This naive parton model was later extended to the theory of Quantum
Chromodynamics (QCD) which provides a justification of the parton picture
altering its predictions by including corrections ensuing from the
quark-gluon interactions.
This interaction is invariant to color SU(3) local gauge transformations.
The success of QCD in the description of hadronic processes is rooted
in the property of asymptotic freedom \cite{Gross:1973id,Politzer:1973fx}
which enables the systematic calculation of short-distance processes
as a series expansion in terms of the effective coupling $\alpha_s(Q^2)$
which vanishes for $Q^2\longrightarrow \infty$ while preserving
renormalization.
Then, one can separate out all binding effects, attributable to
nonperturbative physics, and absorb them into universal parton distribution
functions (PDF)s, parton fragmentation functions (PFF)s, lightcone
distribution amplitudes (DA)s for hadrons, transverse-momentum dependent
(TMD) PDFs (Sec.\ \ref{sec:scimemi}), generalized parton distributions (GPD)s,
(Sec.\ \ref{sec:moutarde}), etc.
Their extraction from lattice computations will be considered in
Sec.\ \ref{sec:alexandrou}, while the particular channels and experiments
to access them by measurements will be addressed in Sec.\ \ref{sec:horn} and
in Sec.\ \ref{sec:EIC} which will present a state of the art on
three-dimensional (3D) nucleon tomography and its future prospects.

\begin{figure}[h]                                                      
\centering
\includegraphics[width=10cm,clip]{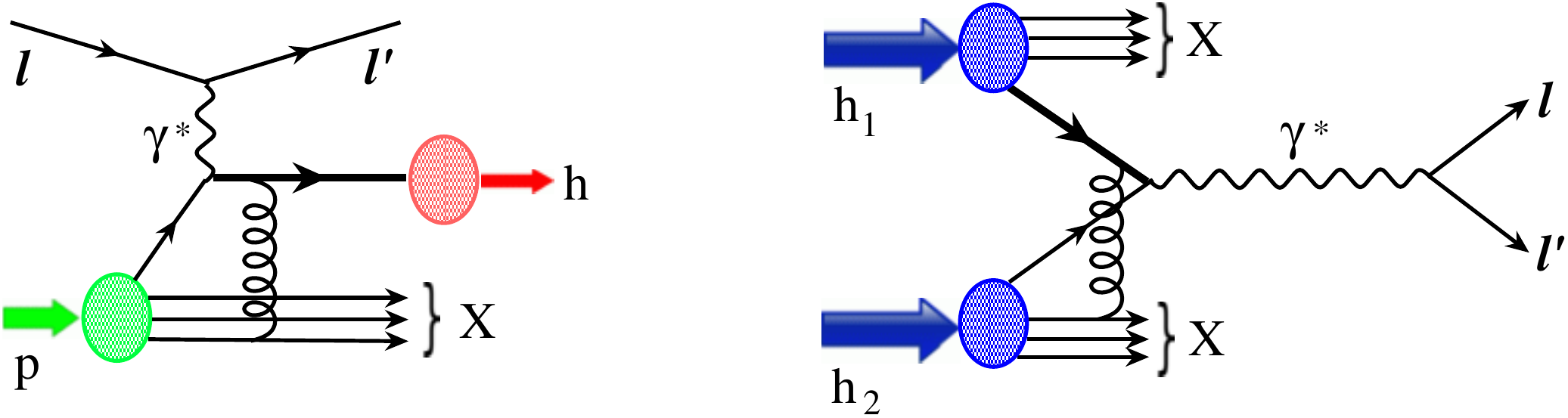} 
\caption{Schematic representation of partonic subprocesses in QCD
``embedded'' within the experimentally measured semi-inclusive deep
inelastic lepton (l) scattering
$l + p \longrightarrow l' + h + X$, where $p$ is the incoming proton
and $h$ represents the detected hadron in the final state (left panel).
The right panel shows the analogous situation for the Drell-Yan process
$h_{1}+h_{2} \longrightarrow \gamma^* + X\longrightarrow l + l' + X$,
where $h_{1(2)}$ represent incoming hadrons.
The thick lines in both panels denote ``eikonalized'', i.e., Wilson-line
extended quarks to account for initial (DY) or final (SIDIS) state
interactions.
Examples of single gluon exchanges emanating from these lines are also shown.
Additional hard-gluon exchanges have been omitted.
In both panels the symbol $X$ represents an inclusive sum over all final states.}
\label{fig:sidis-dy}
\end{figure}

Let us now introduce the theoretical tools to probe the interior structure
of the nucleon in more detail.
To this end, consider, for instance, the longitudinal distribution of partons
inside the nucleon $\text{N}$, entering a deep-inelastic scattering (DIS)
process $l\text{N} \longrightarrow l'X$.
This process, shown for the proton in the left panel of Fig.\ \ref{fig:sidis-dy},
can be expressed in terms of the matrix element\footnote{Boldfaced symbols denote
Euclidean two vectors in the transverse plane.}
\begin{eqnarray}
&& 
f_{q/\text{N}}(x,\mu)
=
  \frac{1}{4\pi} \int_{}^{} dy^- e^{-ix p^+ y^-}
  \langle
         p| \bar{\psi}(0^+,y^-,{\bf{0}}_T)
\gamma^+ \mathcal{W}(0^-,y^-) \psi(0^+,0^-,{\bf{0}}_T) | p
  \rangle
\label{eq:quark-pdf}
\end{eqnarray}
where
$f_{q/\text{N}}(x,\mu)$
is the PDF describing a quark $q$ in a nucleon $N$ carrying a fraction $x$
of its momentum $p$ at the resolution (factorization) scale $\mu$.
Gauge invariance of the correlator is ensured by the insertion of
the Wilson line (or gauge link) operator
\begin{equation}
  \mathcal{W}(0^-,y^-)
=
  P \exp \left[
                ig \int_{0^-}^{y^-} dz^- A_{a}^{+}(0^+, z^-, {\bf{0}}_T)t_a
         \right]
\label{eq:Wilson-line}
\end{equation}
evaluated in the fundamental representation of SU(3) and taken along a
lightlike contour from $0^-$ to $y^-$.
Note that here we are using the lightcone notation
$y^{\pm}=(v^0 \pm y^3)/\sqrt{2}$ for any vector $v^\mu$.

Collinear PDFs, like $f_{a/A}$, where parton $a$ is a quark, antiquark, or
a gluon, in a hadron $A$, represent the universal part of the factorized cross
section of a collinear process, like DIS, and are related to leading-twist
lightcone correlators of electroweak currents in the hadronic tensor
\begin{equation}
  W^{\mu\nu}
=
  \frac{1}{4\pi} \int_{}^{}d^4y e^{iq\cdot y}
  \sum_{X} \left\langle \text{N}|j^\mu (y)|X\right\rangle
  \left\langle X|j^\nu (0)|\text{N}\right\rangle
\label{eq:hadr-tens}
\end{equation}
for the process $l\text{N}\longrightarrow l'X$.
For $Q^2$ large and $x$ fixed, $W^{\mu\nu}$ can be cast in factorized
form (see \cite{Collins:1989gx} and references cited therein) to read
\begin{eqnarray}
W^{\mu\nu}(q^\mu,p^\mu)
=
  \sum_{a} \int_{x}^{1} \frac{d\xi}{\xi}
  f_{a/\text{N}}(\xi, \mu)
H_{a}^{\mu\nu}(q^\mu,\xi p^\mu, \mu, \alpha_s(\mu))
  + \mbox{remainder} \, ,
\label{eq:hadr-fac-the}
\end{eqnarray}
where the contribution of all short-distance subprocesses on the
parton $a$ is denoted by $H_{a}^{\mu\nu}$.
By virtue of universality, the PDFs for the Drell-Yan (DY) process,
shown on the right of Fig.\ \ref{fig:sidis-dy}, should be the same as in
DIS --- left panel of the same figure.
Moreover, the momentum-scale dependence of these PDFs is governed by
the Dokshitzer-Gribov-Lipatov-Altarelli-Parisi (DGLAP)
\cite{Altarelli:1977zs,Gribov:1972ri,Dokshitzer:1977sg}
evolution equation, so that once determined at an initial scale, they can
be evolved in perturbative QCD to any desired reference momentum to
confront theoretical predictions with the experimental data using the
appropriate anomalous dimensions (i.e., splitting functions).
A large set of PDFs has been extracted from global analysis of the
existing data, from the low-momentum to the Large Hadron Collider (LHC)
regime, but this procedure depends on the accuracy of the process-dependent
perturbatively calculated short-distance part $H_{a}^{\mu\nu}$, see
\cite{Brambilla:2014jmp} for a recent review.

Thus, the factorization formalism \cite{Collins:2011zzd} of the $\mu$
dependence contains a strong predictive power for scattering off a nucleon
(hadron).
However, its validity on the partonic level, beyond the collinear
approximation, faces challenges which are related to the appearance of
so-called rapidity divergences ensuing from Wilson lines and their
renormalization (see Sec.\ \ref{sec:scimemi}).
Theoretically, these effects originate from the Wilson-line-extended
structure of the operator definition of quark (gluon) correlators, as
it becomes obvious from the following TMD field correlator
\cite{Ji:2002aa,Belitsky:2002sm,Boer:2003cm}
\begin{eqnarray}
  \Phi_{ij}^{q[C]}(x, \bm{k}_{T};n)
=
  \int_{}^{}\frac{d(y\cdot P)d^2\bm{y}_{T}}{(2\pi)^3}
  e^{ik\cdot y}
  \left\langle
               p| \bar{\psi}_{j}(y)\mathcal{W}(0,y|C)\psi_{i}(0)|p
  \right\rangle_{y\cdot n=0} \, .
\label{eq:TMD-definition}
\end{eqnarray}
One notices the path dependence of this expression encoded in the contour $C$
in the exponential line integral.
It can be resolved by adopting that particular contour which ensures the
continuous color flow in the considered partonic process.
As a result, the DY process, shown in the right panel of Fig.\ \ref{fig:sidis-dy},
contains a sign reversal relative to the SIDIS situation
(left panel in Fig.\ \ref{fig:sidis-dy}), which originates from the change of a
future-pointing Wilson line to one with the opposite orientation as a
consequence of CP invariance and CPT conservation in QCD.
This entails the breakdown of universality, because the factored out
nonperturbative part of the SIDIS setup cannot be used without
readjustment (sign flip) in the DY process:
$
 \left[f_{1Tq}^{\perp}\right]_\text{DY}
=
 -\left[f_{1Tq}^{\perp}\right]_\text{SIDIS}
$ \cite{Collins:2002kn}.
This intriguing behavior constitutes in fact the litmus test of the
TMD approach to single spin asymmetries \cite{Brodsky:2002cx} which
require that the rescattering of the struck quark in the field of the
remnant hadron generates an interaction phase.
This phase would be forced to vanish by the time-reversal invariance
in the absence of the directional dependence of the Wilson line.
Additional phases appear for time-reversal-odd TMD PDFs even at the leading-twist
level when includes into the Wilson lines the Pauli tensor term to account for
a correct treatment of the spin degrees of freedom \cite{Cherednikov:2010uy,Stefanis:2010vu}.

The formal proof of factorization employs a detailed analysis of
singularities that originate from different sources:
(i) ultraviolet (UV) poles, induced by large loop momenta,
that can be regularized dimensionally,
(ii) rapidity divergences that originate from the Wilson lines,
and
(iii) overlapping UV and rapidity divergences.
The latter emerge from gluons moving with an infinite rapidity in the opposite
direction with respect to their parent hadron and cannot be regularized by
infrared gluon mass regulators.
While in the collinear case rapidity divergences cancel in the sum of graphs,
in the TMD case one needs additional regularization parameters.
We will have to say more about these problems in Sec.\ \ref{sec:scimemi} below.
In the same section we will consider the calculation of TMD correlators beyond
the leading order in $\alpha_s(\mu)$ and discuss the concepts of their
evolution.
First attempts to ``measure'' TMD PDFs on the lattice have been given in
\cite{Musch:2011er}, while current investigations will be presented in
Sec.\ \ref{sec:alexandrou}.
A compilation of the various TMD PDFs (``TMDs'' for short) is given in
Table \ref{tab:TMDs}.
Note that a similar structure holds also for gluon TMDs.

\begin{table}[h]
\centering
\caption{Twist-two TMDs as functions of $(x,\bm{k}_T)$
describing correlations between intrinsic spin and transverse
momentum using the following abbreviations: U (unpolarized),
L (longitudinally polarized),
T (transversely polarized).
The boldfaced elements survive the $\bm{k}_T$ integration.
The two terms in brackets are T-odd, whereas all elements in
the last column are chirally odd.}
\label{tab-1}       
\begin{tabular}{|c|c|c|c|}
\multicolumn{4}{c}{Nucleon$\backslash$Quark Polarization} \\ \hline
   & U                        & L                                & T                                                      \\ \hline
 U & $\bm{f_1^q}$             &                                  & $\left[h_{1}^{q\bot}\right]$                           \\
   & unpolarized              &                                  & Boer-Mulders                                           \\ \hline
 L &                          & $\bm{g_{1\text{L}}^{q}}$         & $h_{1\text{L}}^{q\perp}$                               \\
   &                          & helicity                         & worm-gear L                                            \\ \hline
 T & $\left[f_{1\text{T}}^{q\bot}\right]$ & $g_{1\text{T}}^{q\bot}$  & $\bm{h_{1\text{T}}^{q}}$ | $h_{1\text{T}}^{q\bot}$ \\
   & Sivers                   & worm-gear T                      & transversity        | pretzelosity                     \\ \hline
\end{tabular}
\label{tab:TMDs}
\end{table}

\section{Parton distributions with transverse degrees of
freedom --- TMDs}
\label{sec:scimemi}
\footnote{Based on the contribution by I. Scimemi.}The longitudinal PDFs are
based on collinear factorization and provide no
information about the transverse structure of hadrons.
To achieve a 3D picture of the hadronic structure,
one has to retain the transverse momenta $\bm{k}_T$ of the partons
unintegrated, as expressed in Eq.\ (\ref{eq:TMD-definition}).
This gives rise to eight $\bm{k}_T$ dependent PDFs of leading twist two
(see Table \ref{tab:TMDs}),
which enter various processes as the SIDIS and the DY process, both illustrated in
Fig.\ \ref{fig:sidis-dy}.
In SIDIS, one has the convolution of a TMD with a fragmentation function, whereas
in DY one faces the convolution of two TMDs.
In this section, we will discuss the properties of TMDs from the theoretical
point of view and address them in more detail.
On focus is the use of $\bm{k}_T$ factorization theorems, the renormalization of
rapidity singularities, and the TMD evolution behavior in the factorized dynamical
regimes (see \cite{Boer:2011fh} for a comprehensive review).

\begin{figure}[h]                                                      
\centering
\includegraphics[width=8cm,clip]{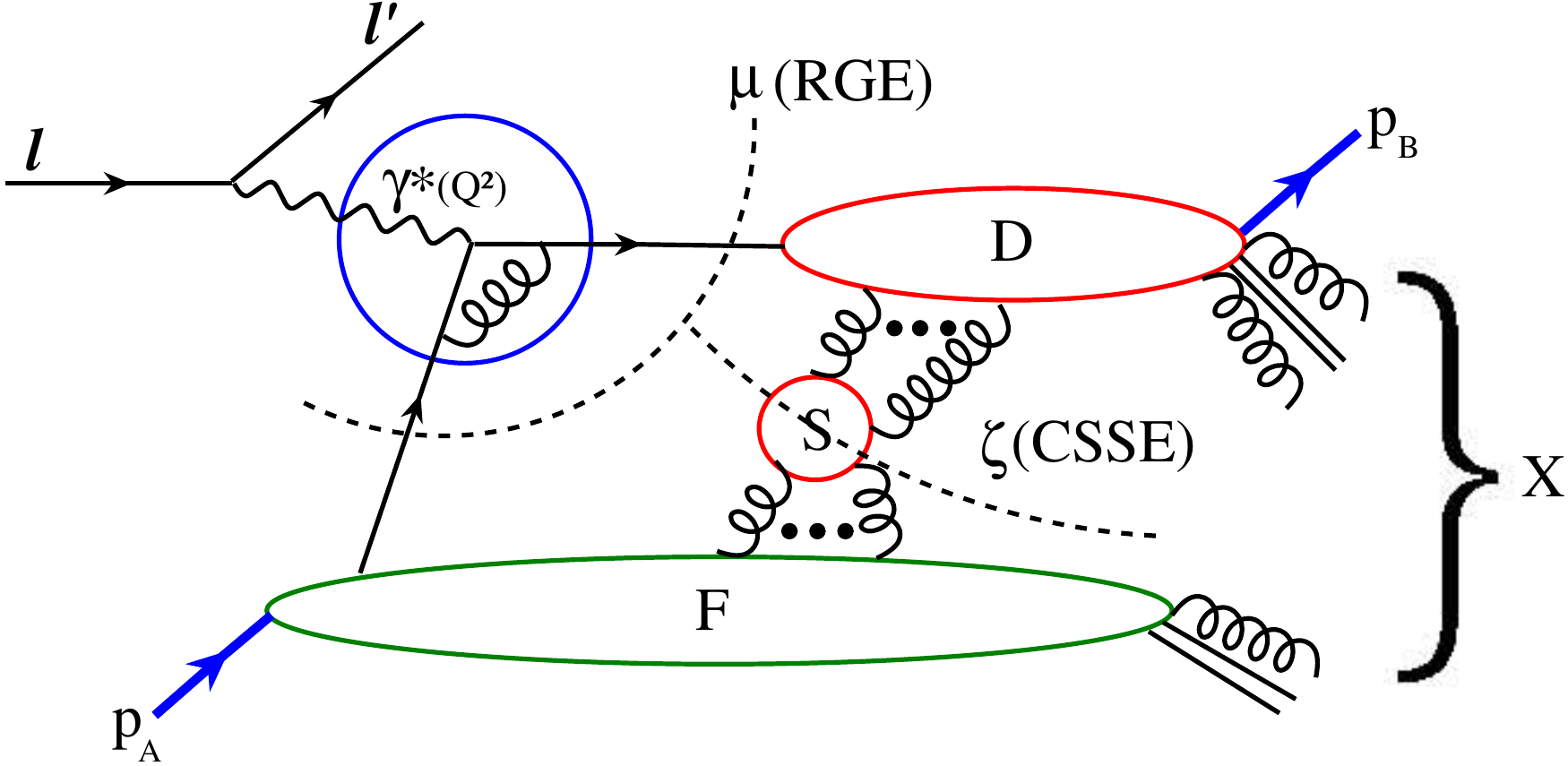} 
\caption{Generic structure of a SIDIS-like process involving various TMD PDFs.
The factorization of dynamical regimes is indicated by dashed lines.
The large $\mu^2$ evolution is controlled by the renormalization-group
equation (RGE), while the evolution with respect to $\zeta$ follows the
Collins-Soper-Sterman (CSS) equation \protect\cite{Collins:1984kg} (see text).}
\label{fig:TMD}
\end{figure}

The core of the TMD factorization theorem in its current form
\cite{Collins:2011zzd,GarciaEchevarria:2011rb,Echevarria:2012js,Echevarria:2014rua},
is based on the understanding of the structure of rapidity divergences in the
DY and/or SIDIS cross section.

As shown in Fig. \ref{fig:TMD}, the cross section for SIDIS can be split,
by virtue of power counting arguments, into three pieces:
a transverse-momentum dependent initial state (called \lq\lq{}F\rq\rq{}),
a final state (termed \lq\lq{}D\rq\rq{}), and a soft-interaction part
(denoted by \lq\lq{}S\rq\rq{}) which connects the previous two.
The power counting procedure, which defines these states, appears naturally in
an effective field-theory framework, like the Soft Collinear Effective Theory
(SCET) \cite{Bauer:2000yr,Bauer:2001yt,Beneke:2002ph}.
However, one can obtain this dissection of the cross section, by employing
more standard QCD arguments in connection with factorization theorems.
Actually, the states so naively identified as above, are not {\it{per se}} well
defined.
This can be checked, for instance, by a one-loop calculation, to show that the
rapidity divergences induce a mixing of all of these states so that it is
impossible to arrive at a rigorous definition of a state whose perturbative
calculation allows the proper separation of ultraviolet and infrared (IR) scales.
In order to achieve this goal, we have to proceed in a different way which is
exposed below.

To this end, we define the {\it bare} (unrenormalized and singular in rapidity) quark,
anti-quark and gluon unpolarized TMD PDF operators as follows:
\begin{align}
\nonumber
O^\text{bare}_q(x,\vecb b_T)&=\frac{1}{2}\sum_X\int \frac{d\xi^-}{2\pi}e^{-ix p^+\xi^-}
\left\{T\[\bar q_i \,\tilde W_n^T\]_{a}\(\frac{\xi}{2}\)
~|X\rangle \gamma^+_{ij}\langle X|~\bar T\[\tilde W_n^{T\dagger}q_j\]_{a}\(-\frac{\xi}{2}\)
\right\},
\\ \nn
O^\text{bare}_{\bar q}(x,\vecb b_T)&=\frac{1}{2}\sum_X\int \frac{d\xi^-}{2\pi}e^{-ix p^+\xi^-}\left\{T\[\tilde W_n^{T\dagger}q_j\]_{a}\(\frac{\xi}{2}\)
~|X\rangle \gamma^+_{ij}\langle X|~\bar T\[\bar q_i \tilde W_n^T \]_{a}\(-\frac{\xi}{2}\)\right\},
\\\label{def_PDF_op}
O^\text{bare}_g(x,\vecb b_T)&=\frac{1}{xp^+}\sum_X\int \frac{d\xi^-}{2\pi}e^{-ix p^+\xi^-}\left\{T\[F_{+\mu} \,\tilde W_n^T\]_{a}\(\frac{\xi}{2}\)|X\rangle \langle X|
\bar T\[\tilde W_n^{T\dagger}F_{+\mu}\]_{a}\(-\frac{\xi}{2}\)\right\},
\end{align}
where $\xi=\{0^+,\xi^-,\vecb b_T\}$ and $n$, $\bn$ are lightcone vectors
($n^2=\bn^2=0,\; n\cdot\bn=2$).
For a generic vector $v$, we have $v^+=\bar n\cd v$ and $v^-=n\cd v$.
Repeated color indices $a$
($a=1,\dots,N_c$ for quarks and $a=1,\dots,N_c^2-1$ for gluons)
are summed up.
The representations of the color SU(3) generators inside the Wilson lines
are the same as the representations of the corresponding partons
(i.e., fundamental representation for quarks and adjoint representation for gluons).
The Wilson lines $\tilde W_n^T(x)$ emerge at the coordinate $x$ and continue to
lightcone infinity along the vector $n$, where they connect to a transverse gauge link
(indicated by the superscript $T$) extending to transverse infinity.

The hadronic matrix elements of the operators, defined in Eq.\ (\ref{def_PDF_op}),
provide the unsubtracted TMDs in accordance with the TMD factorization
theorems \cite{Collins:2011zzd,GarciaEchevarria:2011rb,Echevarria:2012js}:
\begin{eqnarray}\nn
\Phi_{q\leftarrow N}(x,\vecb b_T)&=&\frac{1}{2}\sum_X\int \frac{d\xi^-}{2\pi}e^{-ix p^+\xi^-} \langle N| \left\{T\[\bar q_i \,\tilde
W_n^T\]_{a}\( \frac{\xi}{2} \) |X\rangle \gamma^+_{ij}\langle X|\bar T\[\tilde W_n^{T\dagger}q_j\]_{a}\(-\frac{\xi}{2} \) \right\} | N\rangle,
\\ \nn
\Phi_{\bar q\leftarrow N}(x,\vecb b_T)&=&\frac{1}{2}\sum_X \int \frac{d\xi^-}{2\pi}e^{-ix p^+\xi^-} \langle N|\left\{T\[\tilde
W_n^{T\dagger}q_j\]_{a}\(\frac{\xi}{2} \) |X\rangle \gamma^+_{ij}\langle X|\bar T\[\bar q_i \tilde W_n^T \]_{a}\(-\frac{\xi}{2} \)\right\} |
N\rangle,\nn
\\ \nn
\Phi_{g\leftarrow N}(x,\vecb b_T)&=&\frac{1}{xp^+}\sum_X \int \frac{d\xi^-}{2\pi}e^{-ix p^+\xi^-}
\\&& \times
\langle N| \left\{T\[F_{+\mu} \,\tilde
W_n^T\]_{a}\(\frac{\xi}{2} \)|X\rangle \langle X| \bar T\[\tilde W_n^{T\dagger}F_{+\mu}\]_{a}\( -\frac{\xi}{2} \)\right\} | N\rangle \, ,
\label{def_PDF_opsand}
\end{eqnarray}
where $N$ is a nucleon/hadron.

Here, the variable $x$ represents the momentum fraction carried by a parton
originating from the nucleon (this refers to the TMD labeling rule $f\ot N$).
One notices that at the operator level the TMDs resemble the integrated parton
densities, the only difference being that the parton fields are additionally
separated by the spacelike distance $b_T$.
In order to renormalize correctly the operators and the respective matrix elements,
one has to perform the regularization of the UV, IR {\it and rapidity divergences}.
The UV divergences in the TMDs are removed by the usual renormalization factors.
In order to cancel rapidity divergences, one has to consider both the so-called
zero-bin subtractions and the soft function.
According to the SCET terminology, the ``zero-bin'' represents the soft-overlap
contribution that has to be removed from the collinear matrix element in order to
avoid double counting of soft singularities \cite{Manohar:2006nz}.
The combination of the zero-bin subtraction with the soft function has a very
particular form, which is dictated by the factorization theorem and should be
included in the definition of the TMD operators in the form of a single
``rapidity renormalization factor'' $R$ in order to complete the definition of
the \textit{renormalized} TMD operator.
Then, one has
\begin{eqnarray}\nn
O_{q,\bar q}(x,\vecb b_T,\mu,\zeta)=Z_q(\zeta, \mu)R_q(\zeta,\mu)O^\text{bare}_{q,\bar q}(x,\vecb b_T),
\\\label{renomalized_ops_PDF}
O_{g}(x,\vecb b_T,\mu,\zeta)=Z_g(\zeta, \mu)R_g(\zeta,\mu)O^\text{bare}_{g}(x,\vecb b_T) \, ,
\end{eqnarray}
where $Z_q$ (quark) and $Z_g$ (gluon) are the UV renormalization constants
for the TMD operators and the scale $\zeta$ emerges as a result of splitting
the soft function between the two TMDs $F$ and $D$.

The scales $\mu$ and $\zeta$ are related to the UV and rapidity subtractions,
respectively.
While the UV renormalization factors depend on the UV regularization method and the
regularization scale $\mu$, the ``rapidity renormalization factors'' depend in
addition on the rapidity regularization method and the rapidity scale $\zeta$ as well.
Remarkably, because the soft function is process independent,
the ``rapidity renormalization factors'' turn out to
be process independent as well (see
\cite{Collins:2004nx,Collins:2011zzd,GarciaEchevarria:2011rb,Echevarria:2012js,Echevarria:2014rua}
for general arguments and \cite{Echevarria:2015byo} for an explicit calculation
at the next-to-next-to-leading order (NNLO)).
However, the particular form of the zero-bin subtractions, contained in the factor $R$,
is regulator dependent.
Therefore, one has to fix the order of how to deal with these singular factors exactly.
Following \cite{Echevarria:2016scs}, one can first remove all rapidity
divergences and carry out the zero-bin subtraction, performing subsequently the multiplication
with the $Z$ factors.
In that case, one finds that the factor $R$ contains not only rapidity divergences,
but also explicit UV poles which, however, have already been taken into account by means of
the factor $Z$.
Thus, different subtraction procedures can produce different intermediate expressions,
while the final (UV finite and rapidity-divergences-free) expressions will be the same.

The final definitions for the TMDs entering the SIDIS process in Fig.\ \ref{fig:TMD}
read
\begin{eqnarray}
\label{def:TMD_pdf}
F_{f\ot N}(x,\vecb{b}_T;\mu,\zeta)&=&\langle N|O_f(x,\vecb{b}_T;\mu,\zeta)|N \rangle,\nn
\\\label{def:TMD_ff}
D_{f\to N}(z,\vecb{b}_T;\mu,\zeta)&=&\langle N|^\dagger\mathbb{O}_f(z,\vecb{b}_T;\mu,\zeta)|N \rangle^\dagger \, .
\end{eqnarray}
This definition in conjunction with the TMD factorization theorem implies the following
relation between bare and renormalized TMDs
\begin{eqnarray}
\label{def:TMD=unTMD_pdf}
F_{f\ot N}(x,\vecb{b}_T;\mu,\zeta)&=&Z_f(\mu,\zeta)R_f(\mu,\zeta)\Phi_{f\ot N}(x,\vecb{b}_T),\nn
\\\label{def:TMD=unTMD_ff}
D_{f\to N}(z,\vecb{b}_T;\mu,\zeta)&=&Z_f(\mu,\zeta)R_f(\mu,\zeta)\Delta_{f\to N}(x,\vecb{b}_T) \, .
\end{eqnarray}
The TMD factorization theorem dictates the explicit form of $R_f$ as well.
It is given by the expression
\begin{eqnarray}
R_f(\zeta,\mu)=\frac{\sqrt{S(\vecb{b}_T)}}{\textbf{Zb}} \, ,
\end{eqnarray}
which involves the soft function $S(\vecb{b}_T)$ and the zero-bin contribution
\textbf{Zb},
i.e., the soft overlap of the collinear and the soft sectors entering
the factorization theorem
\cite{Manohar:2006nz,Collins:2011zzd,GarciaEchevarria:2011rb,Echevarria:2012js,Echevarria:2014rua}.

The soft function is defined as the vacuum expectation value of a certain configuration of
Wilson lines pertaining to the process under investigation.
Consider, for instance, the SIDIS process.
Then, one has
\begin{eqnarray} \label{eq:SF_def}
 S(\vecb b_{T})
=
\frac{{\rm Tr}_c}{N_c} \sandwich{0}{\, T\le[S_n^{T\dagger} \tilde S_\bn^T \ri](0^+,0^-,\vecb b_T) \bar
T\le[\tilde S^{T\dagger}_\bn S_n^T\ri](0)}{0} \, .
\end{eqnarray}
The Wilson lines are given by the following ordered exponentials
\begin{eqnarray}
 \label{eq:SF_def2}
S_{n}^T &=& T_{n} S_{n}\,,
\quad\quad\quad\quad
\tilde S_{\bn}^T = \tilde T_{n} \tilde S_{\bn}\,,
\\ \nn
S_n (x) &=& P \exp \left[i g \int_{-\infty}^0 ds\, n \cdot A (x+s n)\right]\,,
\\\nn
T_{\bn} (x) &=& P \exp \left[i g \int_{-\infty}^0 d\tau\, \vec l_\perp \cdot \vec A_{\perp} (0^+,\infty^-,\vec x_\perp+\vec l_\perp \tau)\right]\,,
\\\nn
\tilde S_\bn (x) &=& P\exp\le[-ig\int_{0}^{\infty} ds\, \bn \cdot A(x+\bn s) \ri]\,,
\\\nn
\tilde T_{n} (x) &=& P\exp\le[-ig\int_{0}^{\infty} d\tau\, \vec l_\perp \cdot \vec A_{\perp}(\infty^+,0^-,\vec x_\perp+\vec l_\perp\tau) \ri]\,.
\nn
\end{eqnarray}
The transverse Wilson lines $T_{n}$ are indispensable in singular gauges,
e.g., the lightcone gauge $n \cdot A=0$ (or $\bn \cdot A=0$)
(see \cite{Belitsky:2002sm,Cherednikov:2007tw,Cherednikov:2008ua,Cherednikov:2009wk,Idilbi:2010im,GarciaEchevarria:2011md}),
while in covariant gauges the $T_{n}$'s appear only formally in order
to preserve gauge invariance, but do not contribute.
Pay attention to the fact that the collinear Wilson lines $W_n^T(x)$, used in the TMD operators
given by Eq.\ (\ref{def_PDF_op}), are defined in the same way as the soft Wilson
lines $S_n^T(x)$.
However, we keep them apart because they behave differently under regularization.

A few technical remarks are here in order.
The zero-bin (or overlap) subtraction is a subtle issue and demands particular caution.
(i) The subtraction procedure depends on the regularization method used to remove the rapidity
divergences (see, e.g., \cite{Echevarria:2012js} for a more complete discussion).
(ii) It might be impossible to define the zero-bin (overlap) region in terms of a proper
matrix element for a particular regularization scheme, even if it is calculable.
Choosing a convenient rapidity regularization, the zero-bin subtractions can be related to
a particular combination of soft factors.
Using, for instance, the modified $\delta$-regularization scheme, the zero-bin subtraction
becomes equal to the soft factor: $\textbf{Zb}=S(\vecb{b}_T)$.
In fact, the modified $\delta$-regularization scheme has been employed with the aim to preserve just
this relation, see \cite{Echevarria:2015byo,Echevarria:2016scs}.
A crucial consequence of this relation is that the collinear Wilson lines $W_{n(\bn)}(x)$ and
the soft Wilson lines $S_{n(\bn)}(x)$ assume different regularized forms.
Finally, using the modified $\delta$-regularization, the expression for the rapidity renormalization factor becomes
\begin{eqnarray}\label{reg:R=1/S}
R^f(\zeta,\mu)\bigg|_{\delta\text{-reg.}}=\frac{1}{\sqrt{S(\vecb{b}_T;\zeta)}} \, ,
\end{eqnarray}
which was first explicitly verified at NNLO in \cite{Echevarria:2015usa,Echevarria:2015byo},
and was confirmed for various kinematics in \cite{Echevarria:2016scs}.
By virtue of the process independence of the soft function
\cite{Collins:2011zzd,GarciaEchevarria:2011rb,Echevarria:2012js,Echevarria:2014rua,Collins:2004nx},
the factor $R_f$ is process independent as well.
We complete this discussion by a comment on the TMD formulation used in Ref.~\cite{Collins:2011zzd}.
There, the rapidity divergences are regularized by tilting the Wilson lines off-the-light-cone.
This way, the contributions originating from the overlapping regions and the soft factors can be
recombined in the individual TMDs by properly combining different soft factors with a
partially removed regulator.
This combination entails in our notation the factor $R^f$, i.e.,
\begin{eqnarray}
\label{reg:R=SS/S}
R^f(\zeta,\mu)\bigg|_{JCC}=\sqrt{ \frac{\tilde S(y_n,y_c)}{\tilde S(y_c,y_{\bar n}) \tilde S(y_n,y_{\bar n})}} \, .
\end{eqnarray}
The further steps remain the same as with the $\delta$-regulator technique.

The use of the modified $\delta$-regulator brings within reach a perturbative calculation
at NNLO.
Indeed, the matrix elements for the soft factor have been evaluated at the two-loop order
in \cite{Echevarria:2015byo}.
This has made it possible to calculate all unpolarized TMDs and TMD fragmentation functions (FF)s
at the same order~\cite{Echevarria:2015usa,Echevarria:2016scs}.
In addition, also the calculation of the soft matrix element for double-parton scattering
\cite{Vladimirov:2016qkd} has been carried out at NNLO.
The results obtained for the TMDs confirmed the previous QCD calculations performed in
\cite{Catani:2011kr,Catani:2012qa,Catani:2013tia,Gehrmann:2012ze,Gehrmann:2014yya}.
Employing symmetry arguments for the soft factor, the evolution of all unpolarized TMDs has been
performed at the three-loop order in \cite{Vladimirov:2016dll} finding agreement with a
recent result reported in \cite{Li:2016ctv}.
These calculations use the operator product expansion of the TMDs in the lowest order of the
power expansion which works sufficiently well for asymptotically high transverse momenta.
However, a complete treatment of the TMDs must also include analysis beyond this lowest order,
a subject we expect to become an active field of research in the near future.
For the time being, the nonperturbative structure of the unpolarized TMDs has been modeled
within a renormalon analysis in the limit of high transverse momentum, see \cite{Scimemi:2016ffw}.
Remarkably, it shows a nontrivial entanglement between the transverse momentum and the
Bjorken variables beyond the lowest order in the power expansion.

At the end of this exposition, we turn our attention to the evolution of TMDs.
Recalling Eq.\ (\ref{def:TMD_pdf}), we write
\begin{align}
\mu^2 \frac{d}{d\mu^2}O_f(x,\vecb b_T)=\frac{1}{2}\gamma^f(\mu,\zeta)O_f(x,\vecb b_T),
&&
\label{RGE:mu}
\mu^2 \frac{d}{d\mu^2}\mathbb{O}_f(z,\vecb b_T)=\frac{1}{2}\gamma^f(\mu,\zeta)\mathbb{O}_f(z,\vecb b_T) \, .
\end{align}
The TMD PDF operator and also the TMD FF operator have the same anomalous dimension $\gamma_f$,
which comes solely from the renormalization factor $Z_f$ and is universal on account of
the universality of the hard interactions
\cite{Collins:2011zzd,GarciaEchevarria:2011rb,Echevarria:2012js}.
Applying standard RGE techniques, we obtain
\begin{eqnarray}
\gamma^q(\mu,\zeta)=2\,\widehat{AD}\(Z_2-Z_q\),\qquad \gamma^g(\mu,\zeta)=2\,\widehat{AD}\(Z_3-Z_g\) \, ,
\end{eqnarray}
where $\widehat{AD}$ represents the operator which extracts the anomalous dimension
from the counterterm (i.e., gives the coefficient in front of the leading pole in
$1/\epsilon$ with a $n!$ prefactor, $n$ being the order of the perturbative expansion).
The first term of the perturbative expansion is the cusp anomalous dimension
$\Gamma_\text{cusp}^f$ \cite{Korchemsky:1987wg}, given by
\begin{eqnarray}
\gamma^f=\Gamma_\text{cusp}^f\mathbf{l}_\zeta-\gamma_V^f \, ,
\end{eqnarray}
where we have used the notation
\begin{eqnarray}
\label{def_logarithms}
\mathbf{L}_X\equiv \ln\(\frac{X^2\vecb{b}^2_T}{4e^{-2\gamma_E}}\),~~~
\mathbf{l}_X\equiv \ln\(\frac{\mu^2}{X}\),~~~
\pmb \lambda_\delta\equiv\ln\(\frac{\delta^+}{p^+}\).
\end{eqnarray}
At the level of the renormalization factors, the logarithmic part of the factor
$R_f$ can be unambiguously fixed by means of the relation
\begin{eqnarray}
  \frac{d^2 \ln R_f}{d\ln \mu^2 \, d\ln \zeta}\Bigg|_{f.p}
=
  \widehat{AD}\left[Z_f\(\frac{d \ln R_f}{d\ln \zeta}\)_{s.p}\right]=-\frac{\Gamma^f_\text{cusp}}{2}.
\end{eqnarray}
A similar relation was obtained in \cite{Chiu:2011qc,Chiu:2012ir}.

Most part of the work reported above, has covered unpolarized TMDs.
However, for a complete understanding of the confinement process,
the study of polarization effects and twist expansion of TMDs are
of fundamental importance as well.
Progress has been achieved with respect to the understanding of rapidity divergences
and their treatment via soft factor functions by virtue of the TMD factorization theorem.
The nonperturbative structure of TMDs should also be explored from a theoretical and
an experimental point of view.
Among some recent developments, we like to mention the efforts to combine the TMD
formalism with the jet analysis discussed in \cite{Kang:2016ehg,Kang:2016mcy,Bain:2016rrv}.

\section{3D imaging of the nucleon's partonic content in terms of GPDs} 
\label{sec:moutarde}
\footnote{Based on the contribution by H. Moutarde.}In continuation of the
previous discussion of the spatial distribution
of quarks and gluons, we will describe in this section distributions which
can provide tomographic 3D images of the nucleon
\cite{Burkardt:2000za,Ralston:2001xs}, termed generalized
parton distributions, or (GPD)s for short.
These quantities were initially introduced in connection with the partonic
description of deeply virtual Compton scattering (DVCS) by M\"{u}ller
\cite{Mueller:1998fv}, and independently by Ji \cite{Ji:1996nm},
and Radyushkin \cite{Radyushkin:1997ki}.
More recently, they have also been employed in the description of
deeply virtual meson production (DVMP) \cite{Radyushkin:1996ru,Collins:1996fb}
and in timelike Compton scattering (TCS) \cite{Berger:2001xd}.
A recent review of the use of GPDs can be found in \cite{Mueller:2014hsa} and
broader comprehensive reviews in
\cite{Goeke:2001tz,Diehl:2003ny,Belitsky:2005qn,Boffi:2007yc,Guidal:2013rya},
while global analysis of available data is performed in \cite{Kroll:2012sm}.
The efforts to extract GPDs from experiment will be addressed in
Sec.\ \ref{sec:horn} and Sec.\ \ref{sec:EIC}.

\begin{figure}[h]                                                      
\centering
\includegraphics[width=11cm,clip]{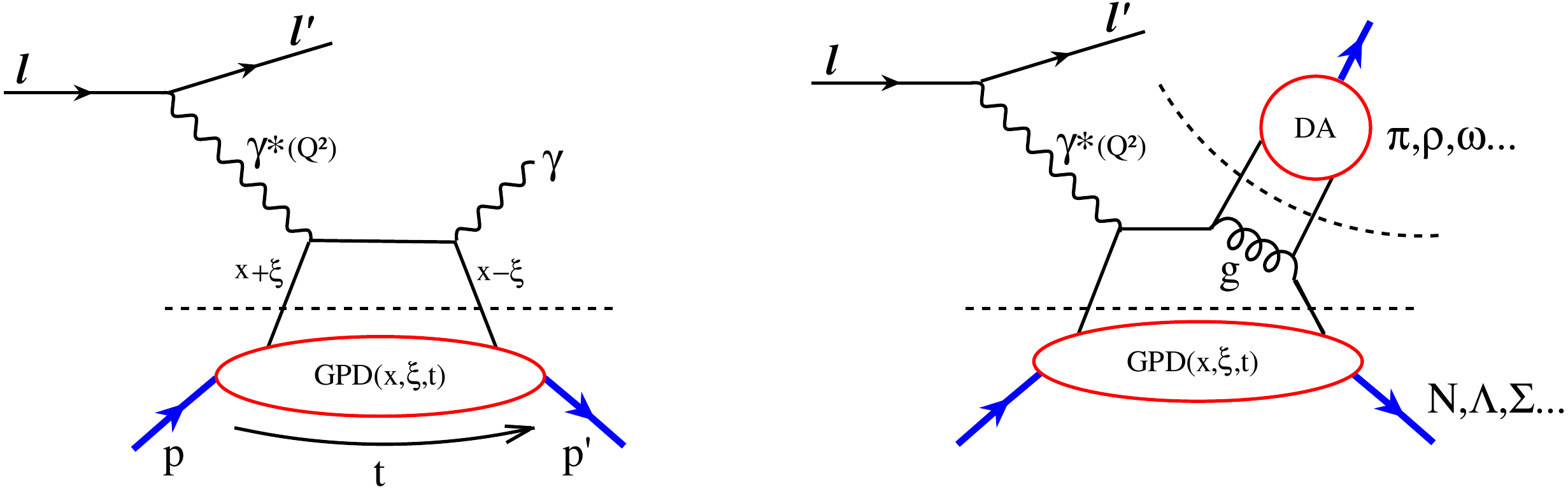} 
\caption{Left panel. Illustration of the DVCS process.
Right panel. Schematic representation of the DVMP.
The factorization of each process is indicated by dashed lines.}
\label{fig:DVCS}
\end{figure}

Unlike PDFs and TMDs, GPDs are defined in terms of non-forward
hadronic matrix elements of quark and gluon correlators, i.e., $p' \neq p$, as
one sees from the generalized form factor for quarks
(see, e.g., \cite{Mueller:2014hsa})
\begin{eqnarray}
  F_{q}(x,\xi,t)
{\hskip -0.07in}
& = &
{\hskip -0.07in}
  \int \frac{dy^-}{2\pi} e^{-i x p^+ y^-}
  \langle p'| \bar{\psi}(\frac{y^-}{2})
          \frac{\gamma^+}{2} \psi(-\frac{y^-}{2}) |p
  \rangle
\\
& \equiv &
{\hskip -0.05in}
  H_q(x,\xi,t) \left[
                     \overline{\cal U}(p')\gamma^\mu{\cal U}(p)
               \right]
                      \frac{n_\mu}{p\cdot n}
+
  E_q(x,\xi,t) \left[
                     \overline{\cal U}(p') \frac{i\sigma^{\mu\nu}(p'-p)_{\nu}}{2M} {\cal U}(p)
               \right]
                      \frac{n_\mu}{p\cdot n} \, ,
\nonumber
\label{eq:quark-gpd}
\end{eqnarray}
where we have suppressed the Wilson lines needed to ensure gauge invariance.
Here, the quantities $H_q(x,\xi,t)$ and $E_q(x,\xi,t)$ are the quark GPDs
which generalize the nucleon form factors $F_1$ and $F_2$ in
Eq.\ (\ref{eq:Dirac-FF}).
They depend on the squared hadron momentum transfer $t=(p' - p)^2$ and
the skewness $\xi=(p' - p)(n/2)$, whereas $x$ is the average longitudinal
momentum fraction carried by the quark.
Note that by replacing in (\ref{eq:quark-gpd})
$\gamma^\mu \longrightarrow \gamma^\mu\gamma_5$,
one obtains two additional quark GPDs, viz.,
$\widetilde{H}_q(x,\xi,t)$ and $\widetilde{E}_q(x,\xi,t)$
so that there are in total four chiral even GPDs defined in transverse space
and longitudinal momentum.
Thus, GPDs provide a snapshot of the nucleon in the $\bm{b}_T$ plane
at each $x$ value.
Analogous expressions are valid also for gluon GPDs.
Employing factorization theorems, the GPDs in DVCS can be linked
to the electroproduction of photons and pseudoscalar/vector
mesons.
At the leading-twist two level, factorization theorems were proved
for transverse polarized photons in DVCS \cite{Collins:1996fb} and
longitudinally polarized photons in DVMP \cite{Collins:1998be}.
These basic hadronic processes involving GPDs are displayed in
Fig.\ \ref{fig:DVCS}.

Evaluating the GPDs for the skewness $\xi\to 0$, one obtains for the
squared hadron momentum transfer
$t\to -\bm{\Delta}_{T}^2$.
This allows one to perform a Fourier transformation of GPD$(x,\xi=0,t)$
with respect to $\bm{\Delta}_T$ and derive adjoint distributions of quarks
and gluons as functions of their longitudinal momentum fraction $x$ and the
transverse position $\bm{b}_T$, $f_a(x,\bm{b}_T)$, where $a=q,\bar{q},g$.
These quantities effectively resemble spatial 3D distributions, i.e.,
tomographic images, of quarks and gluons inside hadrons.
Thus, the combination of GPDs and TMDs can provide a deep and encompassing
3D view on the quark and gluon content of hadrons.

Another key issue in studying the nucleon structure relates to the question of
how the proton spin is distributed among its constituents.
It has been established by measurements of the European Muon Collaboration (EMC)
\cite{Ashman:1987hv} that the proton's quark and antiquark constituents with a given
longitudinal momentum fraction $x$ contribute only about $30~\%$ of its
(longitudinal) spin.
Recent data taken by the RHIC SPIN Collaboration \cite{RHICwp:2012} indicate
that about $15~\%$ of the proton's spin is built up by gluons.
These important discoveries notwithstanding, still the origin of half of
the proton spin is yet unknown.
Because some GPDs are intimately related to the orbital angular momentum carried
by quarks and gluons \cite{Burkardt:2005km}, they are indispensable tools
to reveal the transverse, i.e., transverse-position dependent ($\bm{b}_T$),
structure of the nucleon beyond the collinear approximation, e.g., via DVCS
(see Fig.\ \ref{fig:DVCS}).
An example for the quantification of this connection is provided by Ji's
sum rule \cite{Ji:1996ek}, which expresses the total angular momentum $J_q$
(helicity and orbital momentum) carried by quarks and anti-quarks of the same
flavor, and can be computed in lattice QCD, see \cite{Hagler:2009ni} and
Sec.\ \ref{sec:alexandrou}.

What are the key steps towards nucleon imaging? We first define transverse plane coordinates. Considering a collection of partons confined in a hadron,  labeled by an individual index $i$, flying collinearly at (almost) the speed of light and carrying a longitudinal momentum fraction $x_i$, the transverse center of momentum $R_T$ is defined by
\begin{equation}
\bm{R}_T
= \sum_i x_i \bm{r}_{T i} \, ,
\label{eq:transverse-momentum-center}
\end{equation}
where $\bm{r}_{T i}$ denotes the position of the $i$th parton relative to an arbitrary origin in the transverse plane. Following the conventions of Diehl \cite{Diehl:2003ny}, the 2D transverse plane Fourier transform of a function $f$ evaluated at position $\bm{b}_T$ reads
\begin{equation}
f(\bm{b}_T) = \int \frac{\mathrm{d}^2\bm{D}_T}{(2\pi)^2} e^{- i \bm{D}_T \bm{b}_T} f(t) \, ,
\label{eq:transverse-plane-fourier-transform}
\end{equation}
where the Mandelstam variable $t$ is related to the vector $\bm{D}_T$ through
\begin{equation}
t = t_0 - (1-\xi^2) \bm{D}_{T}^2 \, ,
\label{eq:momentum-transfer-displacement-vector}
\end{equation}
and $t_0 = - 4 \xi^2 m_N^2 / (1-\xi^2)$. Here $|t_0|$ is the smallest value of $|t|$ accessible at given skewness $\xi$. Then, the probability density $\rho_q(x, \bm{b}_T, \lambda, \lambda_N)$ to find a quark $q$ with helicity $\lambda$, carrying longitudinal momentum fraction $x$, at transverse position $\bm{b}_T$, inside a nucleon with longitudinal polarization $\lambda_N$ and transverse spin $S_T$, is \cite{Burkardt:2000za}
\begin{equation}
\rho_q(x, \bm{b}_T, \lambda, \lambda_N) = \frac{1}{2} \left[ H_q(x, 0, \bm{b}_{T}^2) + \frac{\bm{b}_{T}^i \epsilon^{ij} S_{T}^j}{m_N} \frac{\partial E_q}{\partial \bm{b}^2_T}(x, 0, \bm{b}_{T}^2) + \lambda \lambda_N \widetilde{H}_q(x, 0, \bm{b}_{T}^2) \right] \, .
\label{eq:transverse-plane-probability-density}
\end{equation}
From symmetry considerations, the Fourier transform of a generic GPD $F_q$ can only depend on the modulus of $\bm{b}_T$, which is made transparent by the reduction of Eq.\ (\ref{eq:transverse-plane-fourier-transform}) to a one-dimensional integral
\begin{equation}
F_q(x, 0, \bm{b}_{T}^2) = \int_0^{+\infty} \frac{\mathrm{d}|\bm{D}_{T}|}{2\pi} |\bm{D}_{T}| J_0(|\bm{b}_{T}| |\bm{D}_{T}|) F_q(x, 0, -\bm{D}_{T}^2) \, ,
\label{eq:hankel-transform-density}
\end{equation}
where $J_0$ is the 0th-order Bessel function. Therefore the 3D view of the nucleon that we may hope to extract from GPDs requires the knowledge of the three GPDs $H$, $E$ and $\widetilde{H}$, the last two GPDs inducing the deviation from rotational invariance in the transverse plane.

Therefore, the path to nucleon tomography through GPDs is the following ($F$ being a generic notation for a GPD):
\begin{enumerate}
\item \label{item-gpd-extraction} Extract $F(x, \xi, t)$ from experimental data.
\item \label{item-extrapolation-vanishing-skewness} Extrapolate $F(x, \xi, t)$ to vanishing skewness $F(x, 0, t)$.
\item \label{item-extrapolation-infinite-t} Extrapolate $F(x, 0, t)$ up to infinite $t$.
\item \label{item-fourier-transform-computation} Compute the 2D transverse plane Fourier transform of $F(x, 0, t)$ using Eq. (\ref{eq:hankel-transform-density}).
\item \label{item-uncertainty-propagation} Propagate statistical and systematic experimental uncertainties, and theoretical uncertainties through the whole computing chain.
\item \label{item-extrpolation-control} Control extrapolations and evaluate the corresponding uncertainties.
\end{enumerate}
Task \ref{item-gpd-extraction} has been recently reviewed in \cite{Kumericki:2016ehc}, where the current status is described, as well as the prospects offered by future data from Jefferson Lab, COMPASS or EIC. This problem is much harder than PDF fitting, notably because of the \emph{curse of dimensionality} (more functions depending on more variables), but the first decade of DVCS or DVMP fits show the feasibility of GPD extractions from experimental data. It should nevertheless be stressed that global fits are needed, in particular to separate the contributions from gluons (HERA, EIC), sea quarks (HERMES, COMPASS) and valence quarks (CLAS, Hall A).

Task \ref{item-extrapolation-vanishing-skewness} may turn out to be easier than other well-known extrapolations in hadronic physics, as, e.g., proton radius extractions. In the latter, one has to evaluate the derivative of the form factor $F_1$ at vanishing momentum transfer $t$, while experimental data are collected only for non-vanishing momentum transfer. In the former, experimental data exist only for non-vanishing $\xi$, and the available physical range is bounded by kinematic considerations. We can hope at best to extract data for $\xi \in [\xi_{\textrm{min}}, \xi_\textrm{max}]$ with $0 < \xi_{\textrm{min}} <  \xi_\textrm{max} < 1$. Let us now assume that we know a GPD $F$ over this interval. The well-known polynomiality condition asserts that the $n$th Mellin moment of this GPD is a polynomial in the variable $\xi$. Knowing a polynomial on a given interval is enough to know it everywhere in the complex plane, and in particular at $\xi = 0$. Thus, from measurements at $\xi \neq 0$, we may hope to get the Mellin moments of GPDs at $\xi = 0$. Moreover, the support condition $x \in [-1, +1]$ ensures that this type of moment problem, called Hausdorff moment problem, admits a unique solution. Thus, it should be possible to recover the GPD at vanishing skewness $F(x, \xi=0, t)$ without any bias related to the choice of an extrapolation formula.

Task \ref{item-extrapolation-infinite-t} is less constrained. Some model dependence will probably be unavoidable even if the $t$-behavior of the GPDs should follow the form-factor sum rules linking the integral over $x$ of GPDs to form factors. An essential theoretical progress was achieved in 2012 through the computation of finite-$t$ and target-mass corrections to DVCS \cite{Braun:2012hq}. Factorization theorems \cite{Collins:1996fb, Collins:1998be} give a partonic interpretation of DVCS or DVMP under the assumption of the presence of one (unique) large scale in the process, usually the virtuality $Q^2$ of the exchanged photon (see Fig. \ref{fig:DVCS}). In the case of DVCS, this means that the nucleon mass $m_N$ and the momentum transfer $t$ are small compared to $Q^2$. Unfortunately, most of the experimental data collected so far possess a large $Q^2$, with $|t|/Q^2$ ranging from $\simeq 0.04$ in HERA to $0.15$ in CLAS. A significant part of them may be \textit{a priori} excluded from a fit relying on a conservative use of factorization theorems as, e.g., $|t|/Q^2 < 0.1$. Therefore, the state-of-the-art expressions of the DVCS amplitude including these $|t|/Q^2$ and $m_N^2/Q^2$ corrections will play a key role in the analysis of future data by extending the $t$-range that can be used to actually constrain the GPD shapes from experimental data.

After completion of the first three tasks, the computation of the Fourier transform becomes a purely numerical problem. The challenge will be the evaluation of the corresponding integrals with a $\simeq 0.1\%$ systematic uncertainty. Future Jefferson Lab data indeed promise a statistical uncertainty at the level of a few percent. Evaluating the integral of Eq. (\ref{eq:hankel-transform-density}) with a similar uncertainty would make no sense since it would spoil the accuracy of the experimental data in the imaging process. Once all physical hypotheses are stated, the numerical evaluation of an integral should be obtained with a precision at least an order of magnitude smaller than that of measurements. It is reasonable to expect such a control of the numerics in nucleon tomography because it has been achieved, e.g., in the computation of Compton Form Factors (CFF) \cite{Moutarde:2013qs} --- the coefficient functions of the DVCS process.

The propagation of experimental uncertainties of task \ref{item-uncertainty-propagation} can be handled by repeating the previous steps using Monte Carlo replicas of the experimental data generated with appropriate probability densities. Such a program was explicitly carried on in the case of Compton Form Factor fits in the valence region \cite{Guidal:2013rya}. This part of the problem is therefore limited by computing resources, not by some physical principles. Part of the theoretical uncertainties (order of perturbative expansion in the coefficient functions or in GPD evolution equations, relation between the factorization scale and the virtuality $Q^2$ of the exchanged photon, approximations in the expressions of DVCS or DVMP cross sections, etc.) can be systematically probed by means of a flexible framework to perform GPD computations. Such a framework is under construction but will be released to the whole community as soon as its testing period is over. It's the PARTONS\footnote{PARtonic Tomography Of Nucleon Software} project and its software architecture was described in \cite{Berthou:2015oaw}. To state things simply, the PARTONS framework allows systematic differential studies through a simple interface allowing physicists to change models, assumptions or parameterizations to perform the same computations, and evaluate the dispersion of results.

The remaining task \ref{item-extrpolation-control} addresses the question of GPD parameterizations. Today, there is no known parameterization of GPD relying only on first principles, and some control of the extrapolations to large $t$ and small $\xi$ will be needed. As said before, the key to the small $\xi$ extrapolation is probably the polynomiality condition, which is an expression of Lorentz covariance. It goes without saying, that nucleon tomography can possibly be achieved only with models embedding a decent implementation of Lorentz covariance. A promising venue to obtain models fulfilling the polynomiality and positivity properties consists in the description of a GPD as an overlap of light front wave functions in the DGLAP region (defined by $|\xi| < |x|$) and its covariant extension to the complementary ERBL region (defined by $|x| < |\xi|$). A contemporary status on this model building program can be found in \cite{Mezrag:2016hnp}. At last, neural networks may play an important role in the propagation of errors through extrapolations. They are commonly used in PDF fits, and were used in CFF fits in a pioneering study in 2011 \cite{Kumericki:2011rz}.

To summarize, nucleon imaging through GPDs has witnessed tremendous progress over the last decade, showing the maturity of the field from the phenomenological point of view. Nucleon imaging is difficult, but there are good reasons to think that most of the conceptual problems are solved, or close to being solved. If the awaited experimental data are as accurate as expected, and if an EIC is built to explore in detail the gluon sector, then precision tomography of the nucleon is within reach
(see Sec.\ \ref{sec:EIC}).

\section{Nucleon structure on the lattice --- novel results} 
\label{sec:alexandrou}
\begin{figure}[h]                                                      
\centering\vspace*{5mm}
\sidecaption
\includegraphics[width=7cm,clip]{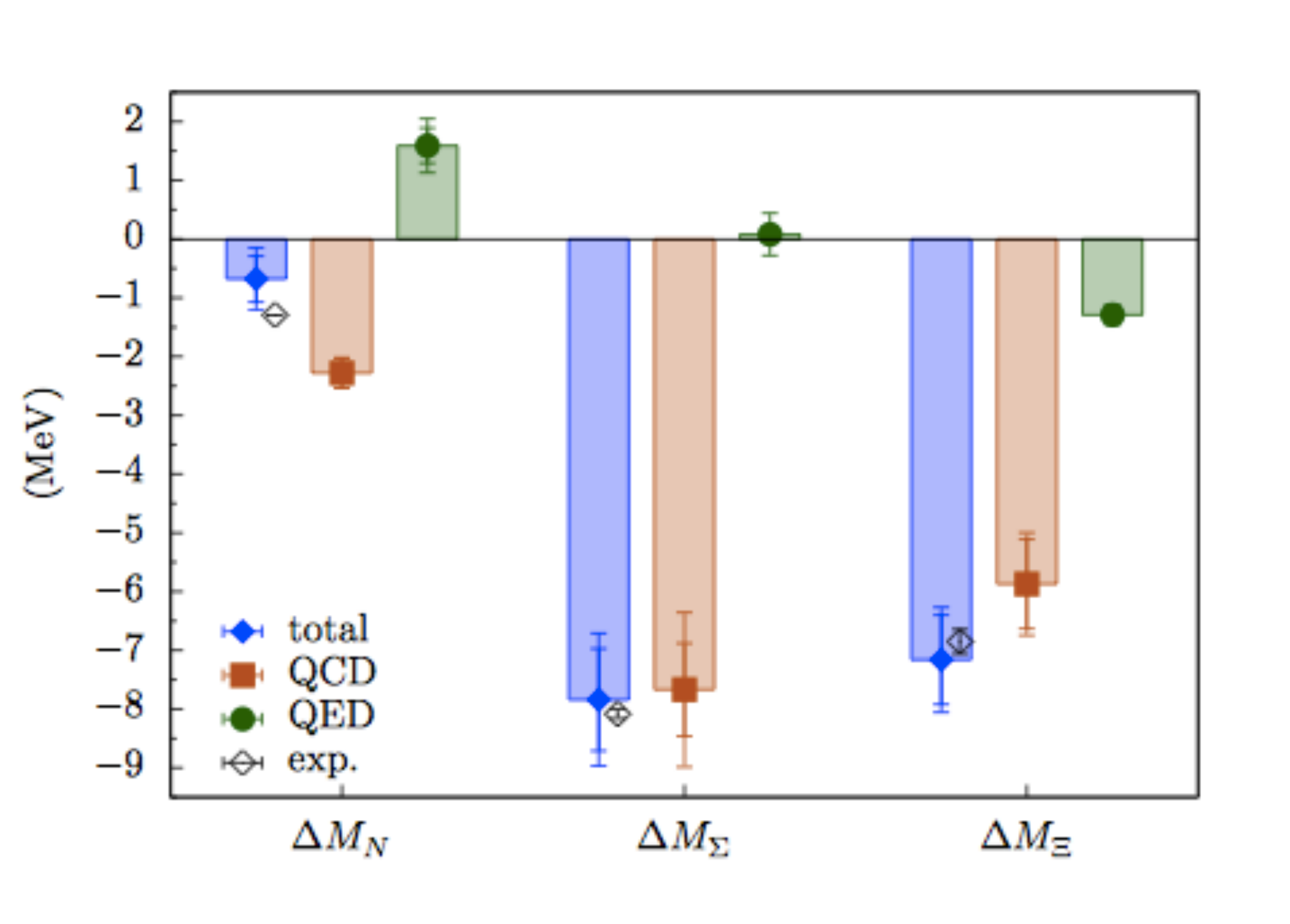} 
\caption{Isospin mass splitting obtained by the BMW Collaboration
in \cite{Borsanyi:2014jba}.
The isospin and electromagnetic effects enter the nucleon mass
with opposite signs.
The physical mass splittings are reproduced.}
\label{fig:BMW}
\end{figure}

\footnote{Based on the contribution by C. Alexandrou.}
Lattice QCD calculations start directly from the QCD Lagrangian defining
it on a four-dimensional discretized hyper-cube proposed by K.~Wilson
\cite{Wilson:1974sk} and utilizing as input the bare quark masses and the
coupling constant or equivalently the lattice spacing.
State-of-the-art simulations have seen a tremendous progress due to both
better formulations and algorithms as well as more powerful computers.
Nowadays we simulate the full QCD near the physical values of the
parameters and reproduce key hadronic properties such as the values of
the low-lying hadron masses including isospin breaking effects due to
the difference in the up and down quark masses and electromagnetism as
shown in Fig.~\ref{fig:BMW}.

\begin{figure}[h!]                                                      
\centering\vspace*{1mm}
\includegraphics[width=0.295\textwidth]{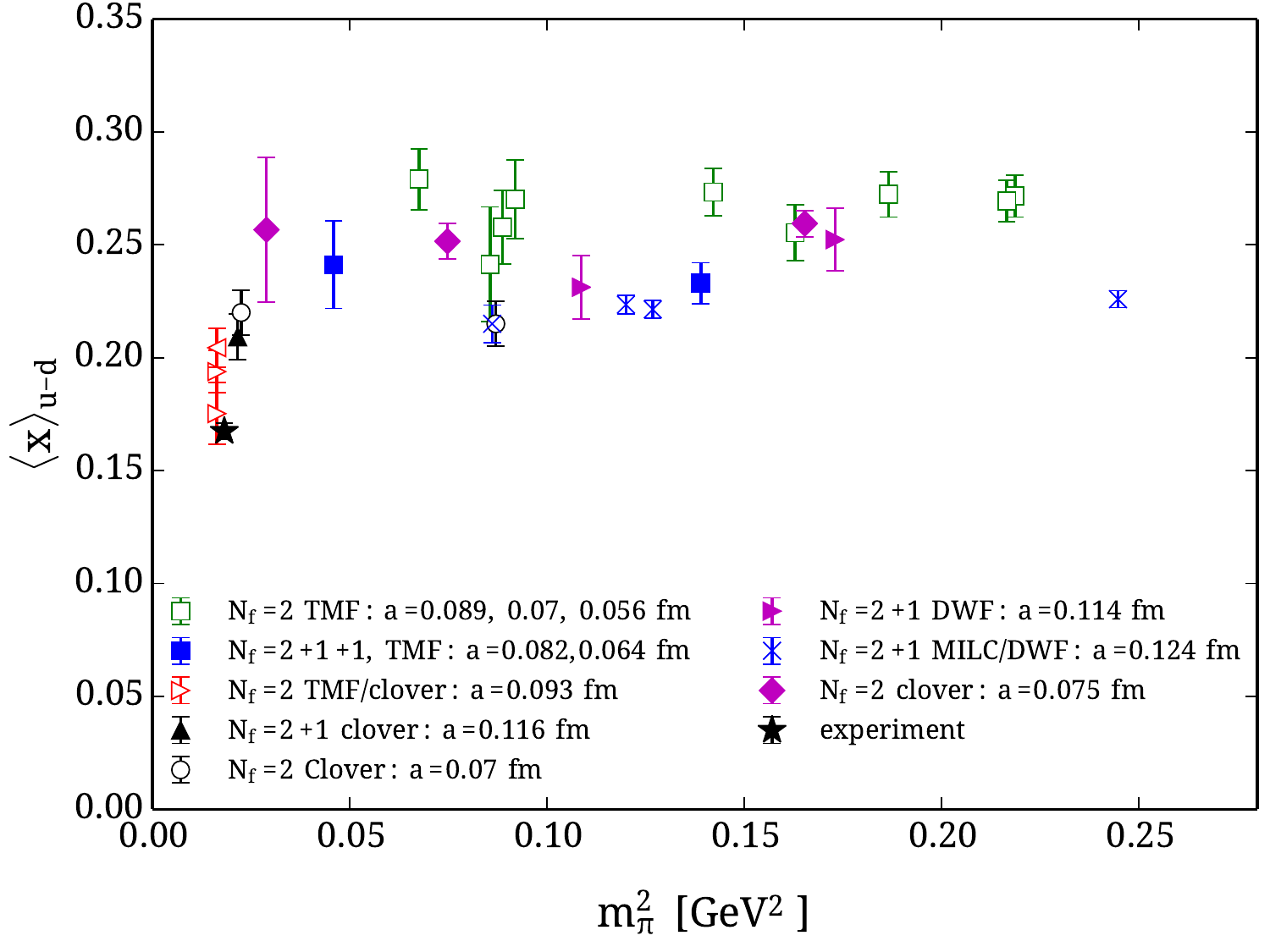} 
\hfill
\includegraphics[width=0.33\textwidth]{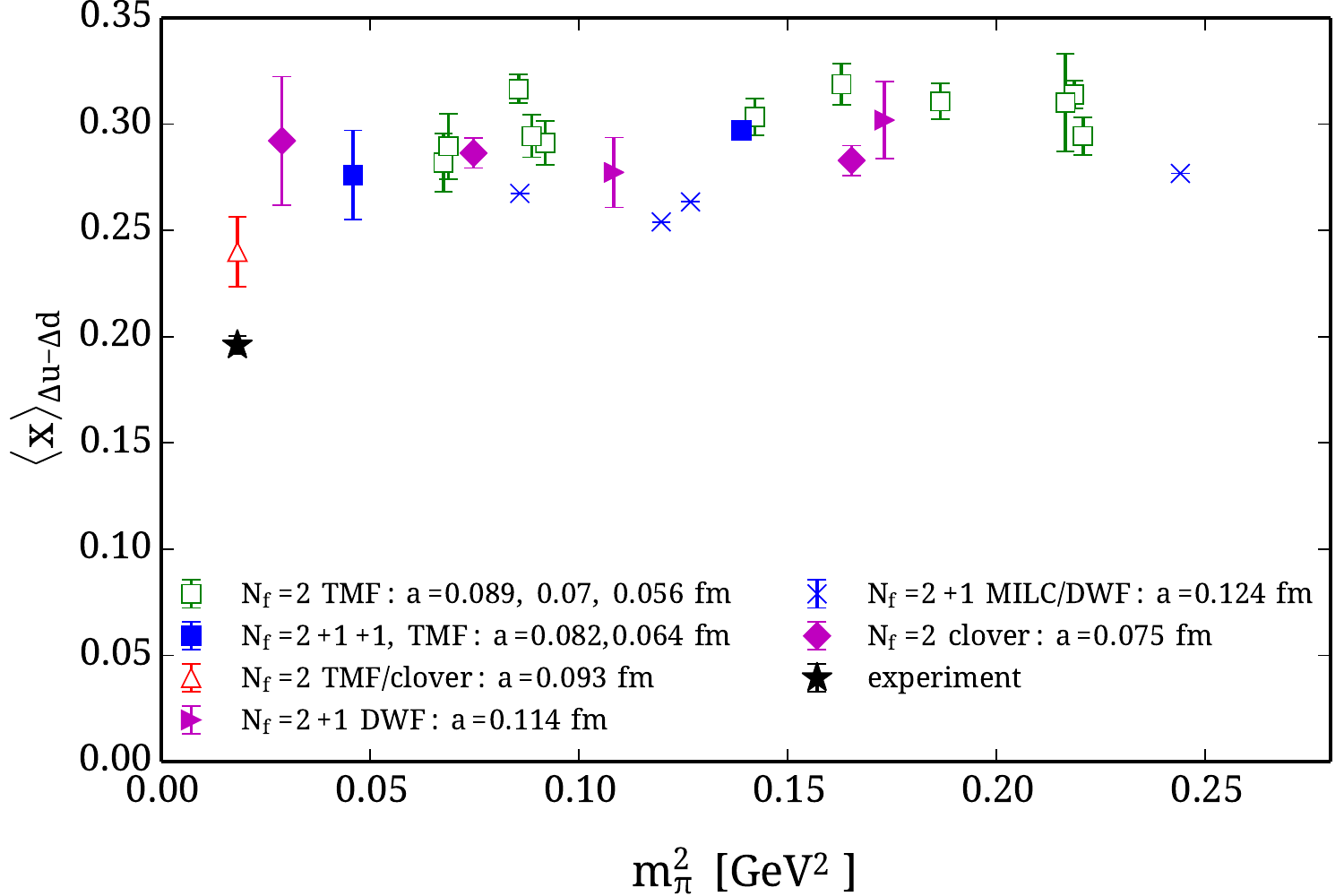} 
\hfill
\includegraphics[width=0.34\textwidth]{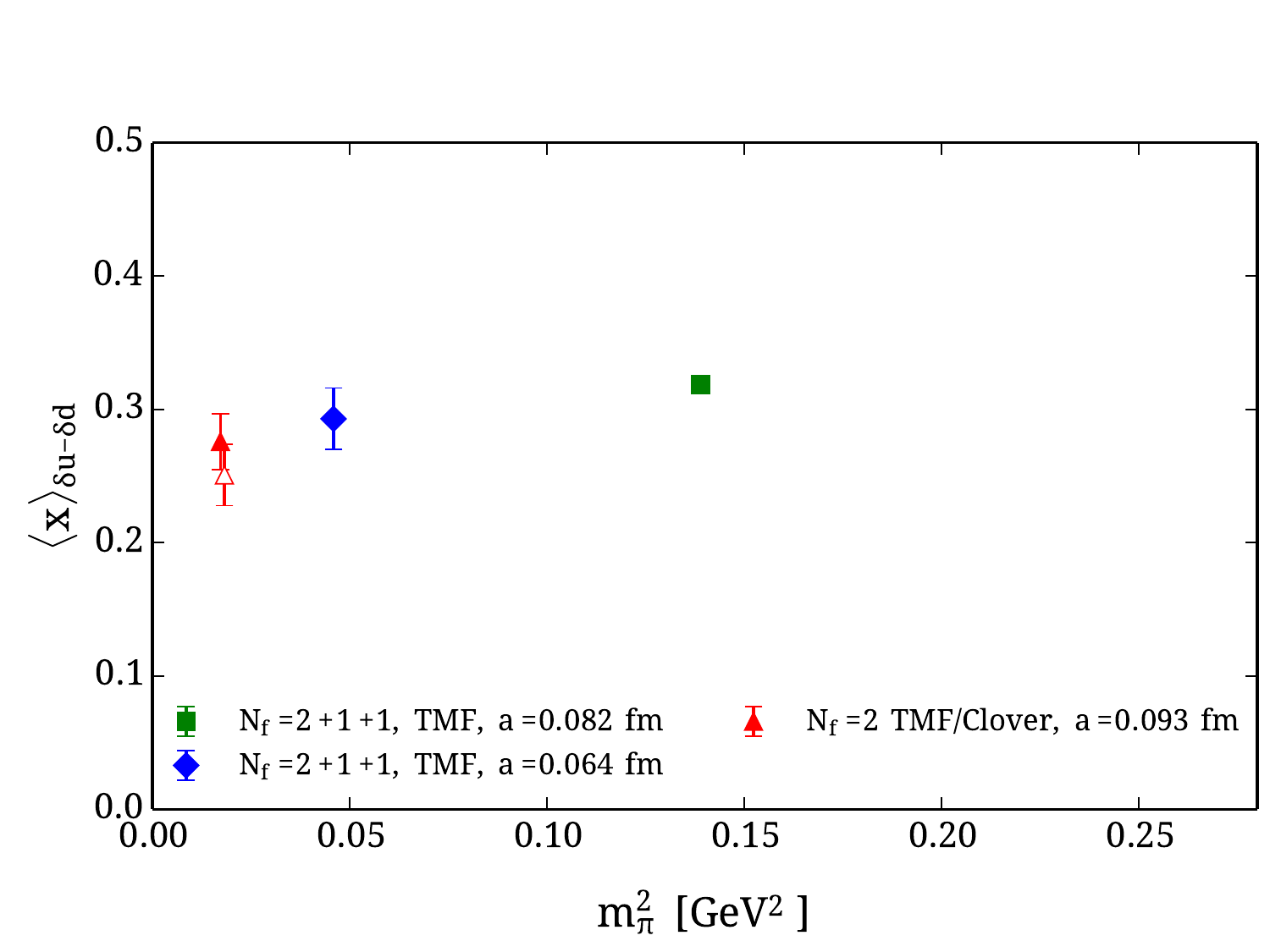} 
\caption{Lattice QCD calculations for the isovector
quark momentum fraction
$\langle x \rangle_{u-d}$ (left),
helicity
$\langle x \rangle_{\Delta u-\Delta d}$ (middle)
using various discretization schemes.
The right panel shows twisted-mass results for transversity
\cite{Abdel-Rehim:2015owa}.
Experimental values are marked by asterisks.}
\label{fig:moments}
\end{figure}

Moments of GPDs have also been computed in lattice QCD providing
insight into the distribution of spin, momentum fraction, etc., as well as
observables that give hints for physical phenomena beyond the physics of
the standard model, for example, scalar and tensor interactions.
An example of the successful application of lattice QCD is the computation
of the first moments of GPDs.
In Fig.~\ref{fig:moments} we show the isovector quark momentum fraction,
helicity and transversity.
For the former two quantities, lattice QCD provides a postdiction, while
for the latter the obtained result constitutes a genuine prediction.

The generalized form factors
$A_{20}(Q^2)|_{Q^{2}=0}$ and $B_{20}(Q^2)|_{Q^{2}=0}$,
where $Q^2=-q^2$ is the momentum transfer squared,
are extracted from the nucleon matrix element of the vector operator
containing one derivative and provide valuable information on the
proton spin.
We have computed
$A^q_{20}(0)=\langle x \rangle_q$
and $B^q_{20}(0)$ for $q=u,d,s$, and also
$A^g_{20}(0)=\langle x \rangle_g$ for the gluon.
We find
\begin{equation}
  \sum_{q=u,d,s}\langle x \rangle_q
=
  0.74(10)\, \,
  \langle x \rangle_g
=
  0.27(2)(2)
\end{equation}
in the $\overline{\rm MS}$ at 2~GeV, where the mixing of
$\langle x\rangle_{u+d+s}$
with the gluon operator is perturbatively computed using one-loop
lattice perturbation theory~\cite{Alexandrou:2016ekb}.
The systematic error results from the difference between using
one- and two-levels of stout smearing.
Having obtained both the quark and the gluon momentum fractions,
we can check the momentum sum
$
 \sum_q \langle x \rangle_q+
        \langle x \rangle_g
=
 1.01(10)(2)
$,
which is very well satisfied.
The proton can be written as
$
 J_N
=
 \frac{1}{2}\sum_q
                  \left(A^q_{20}+B^q_{20}(0)\right)
                + \left(A^g_{20}+B^g_{20}(0)\right)
$.
We have found that $B^q_{20}$ is consistent with zero.
Assuming $B^g_{20}(0)\sim 0$, we can check the spin sum.
We get
$J^{u+d+s}=0.374(51)(42)$
which, when combined with the gluon contribution of
$0.136(12)(12)$,
yields $J_N=0.51(5)(4)$,
in accord with the spin sum of $1/2$ \cite{Alexandrou:2016tuo}.

A new approach that could allow us to obtain the parton distribution
functions directly from lattice QCD has been proposed by
Ji \cite{Ji:2013dva}.
One considers the matrix element
\begin{equation}
  \tilde{q}(x,\Lambda,P_3)
=
  \int_{-\infty}^{+\infty} \frac{dz}{4\pi}
                           e^{-izxP_3}{\langle P|
                           \bar{\psi}(z,0)\,\gamma_3 \,\mathcal{W}(0\to z)\psi(0,0)
                                       |P\rangle}_{h(P_3,z)} \, ,
\end{equation}
where $\tilde{q}(x)$ is the quasi-distribution to be related to the PDFs,
$\Lambda$ is a UV cut-off scale like $1/a$, with $a$ being the
lattice spacing, and $P_3$ is the nucleon's momentum in the $z$ direction.
The above correlator includes a Wilson line $\mathcal{W}(0\to z)$ extending from the
location of the quark field at 0 to $\infty$ along the $z$ direction \cite{Ji:2013dva}.

\begin{figure}[h!]                                                      
\centering\vspace*{1mm}
\includegraphics[width=0.32\textwidth]{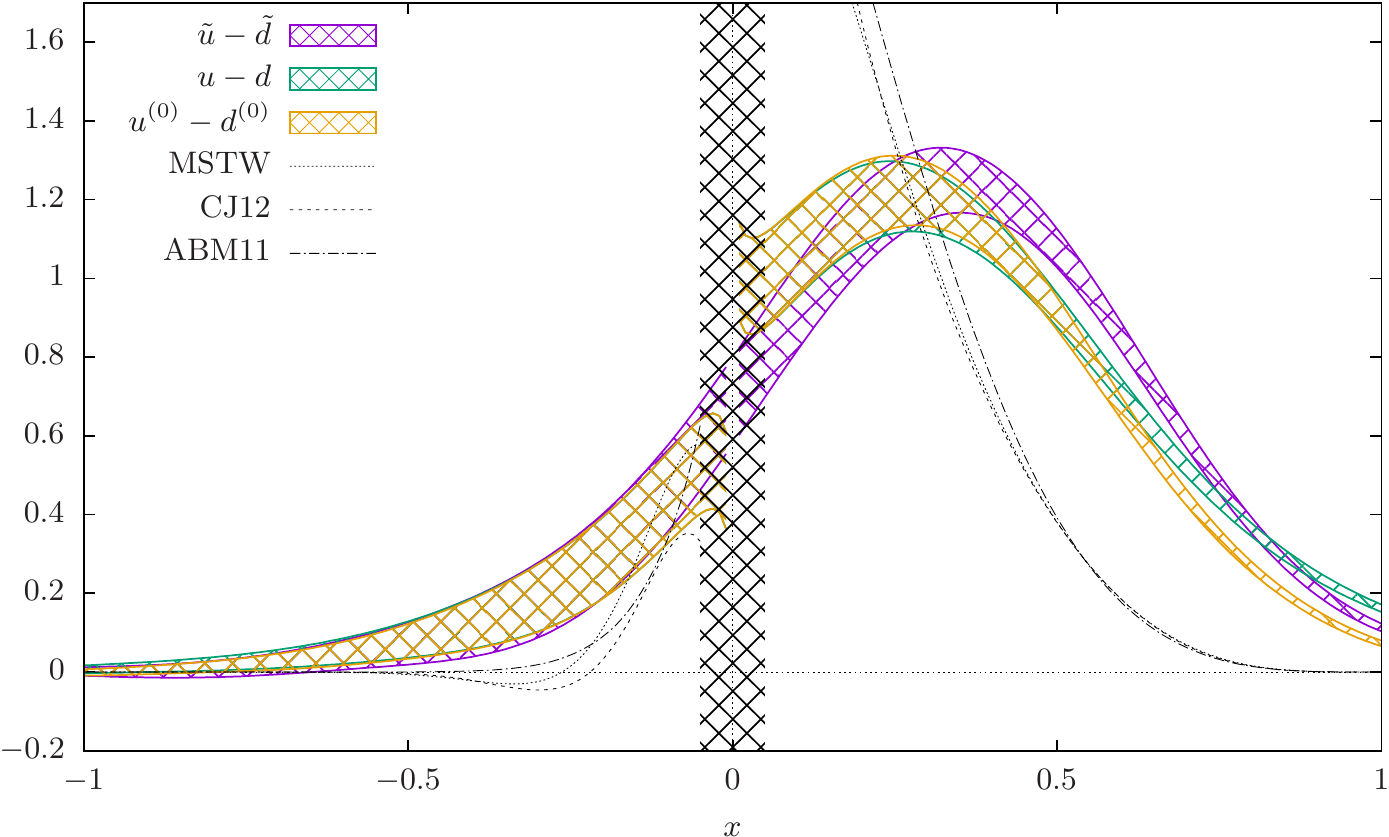} 
\hfill
\includegraphics[width=0.328\textwidth]{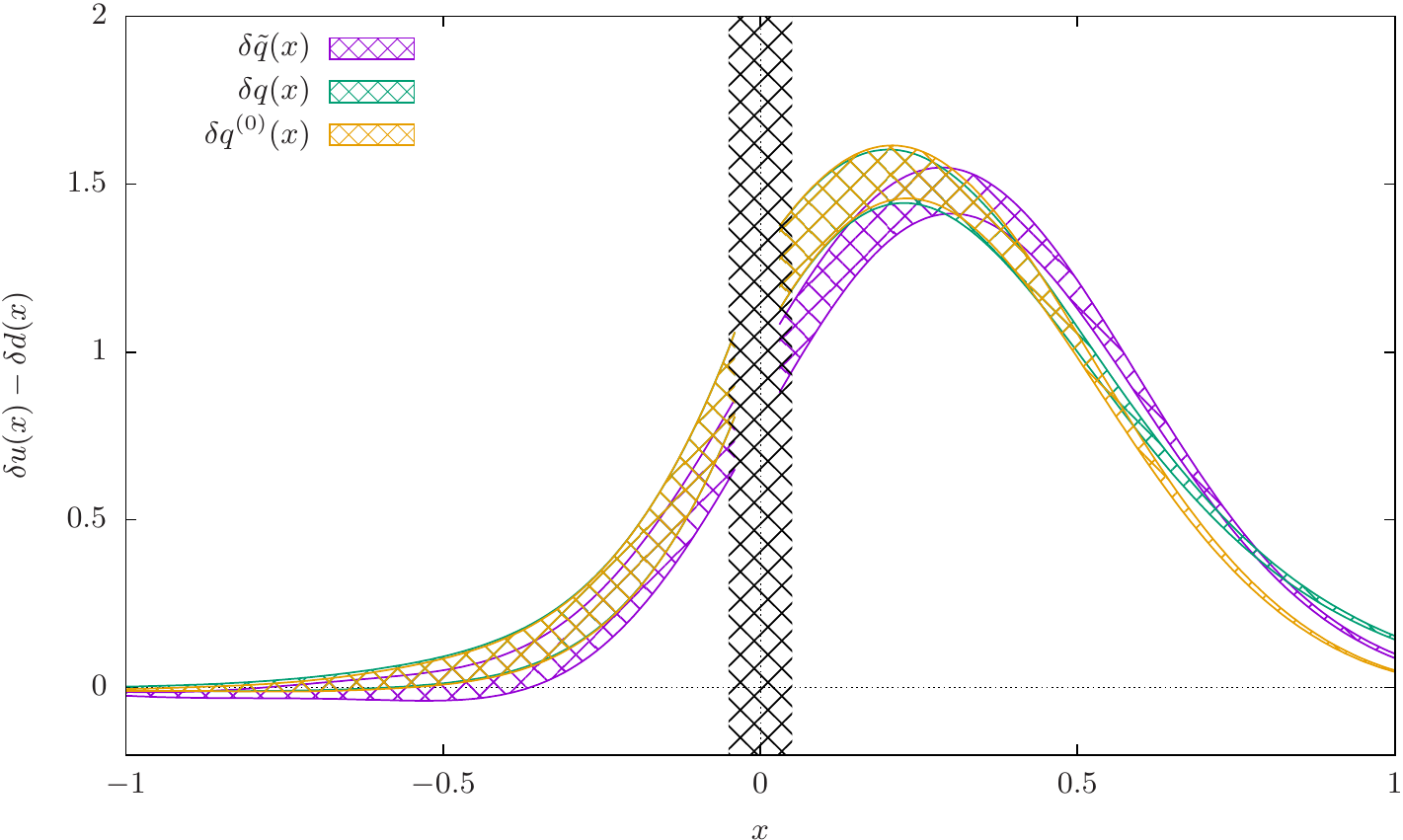} 
\hfill
\includegraphics[width=0.32\textwidth]{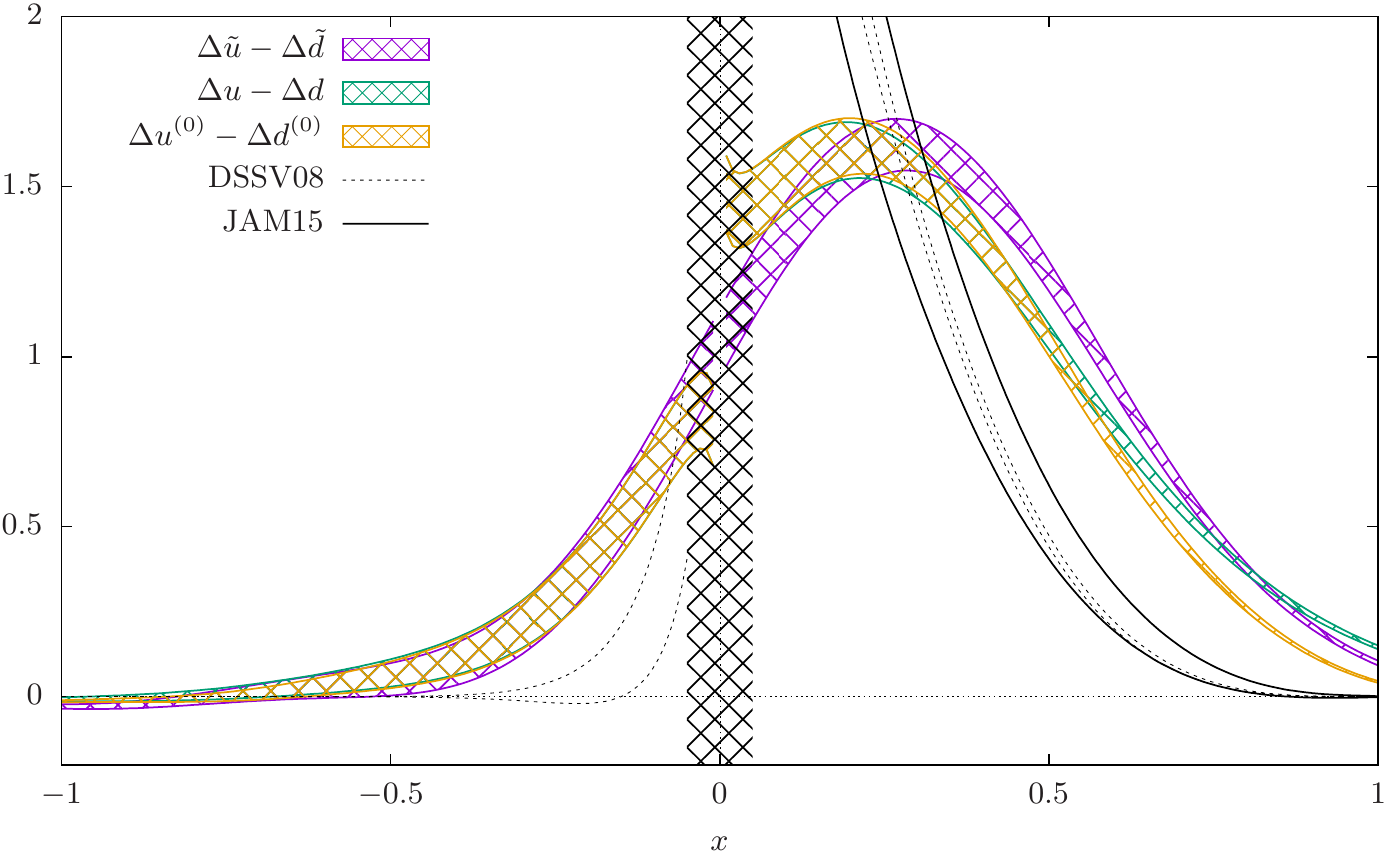} 
\caption{The isovector unpolarized, polarized, and transversity
parton distribution
(figure taken from Ref.~\cite{Alexandrou:2016jqi}).}
\label{fig:PDF}
\end{figure}

First results have been obtained for $N_f=2+1+1$ (two degenerate u and d quarks,
plus a strange and a charm quark) clover fermions on a HISQ
sea~\cite{Lin:2014zya,Chen:2016utp}, and for  $N_f=2+1+1$ twisted-mass
fermions~\cite{Alexandrou:2015rja,Alexandrou:2016jqi} using simulations with
a pion mass of about 310~MeV and 370~MeV, respectively.
In  Fig.\ \ref{fig:PDF} we show the unrenormalized PDFs obtained with the
twisted-mass fermions.
Efforts are under way to understand the renormalization of these quantities
in more detail.

\begin{figure}[h!]                                                      
{\begin{minipage}{0.55\linewidth}
\includegraphics[width=0.77\linewidth]{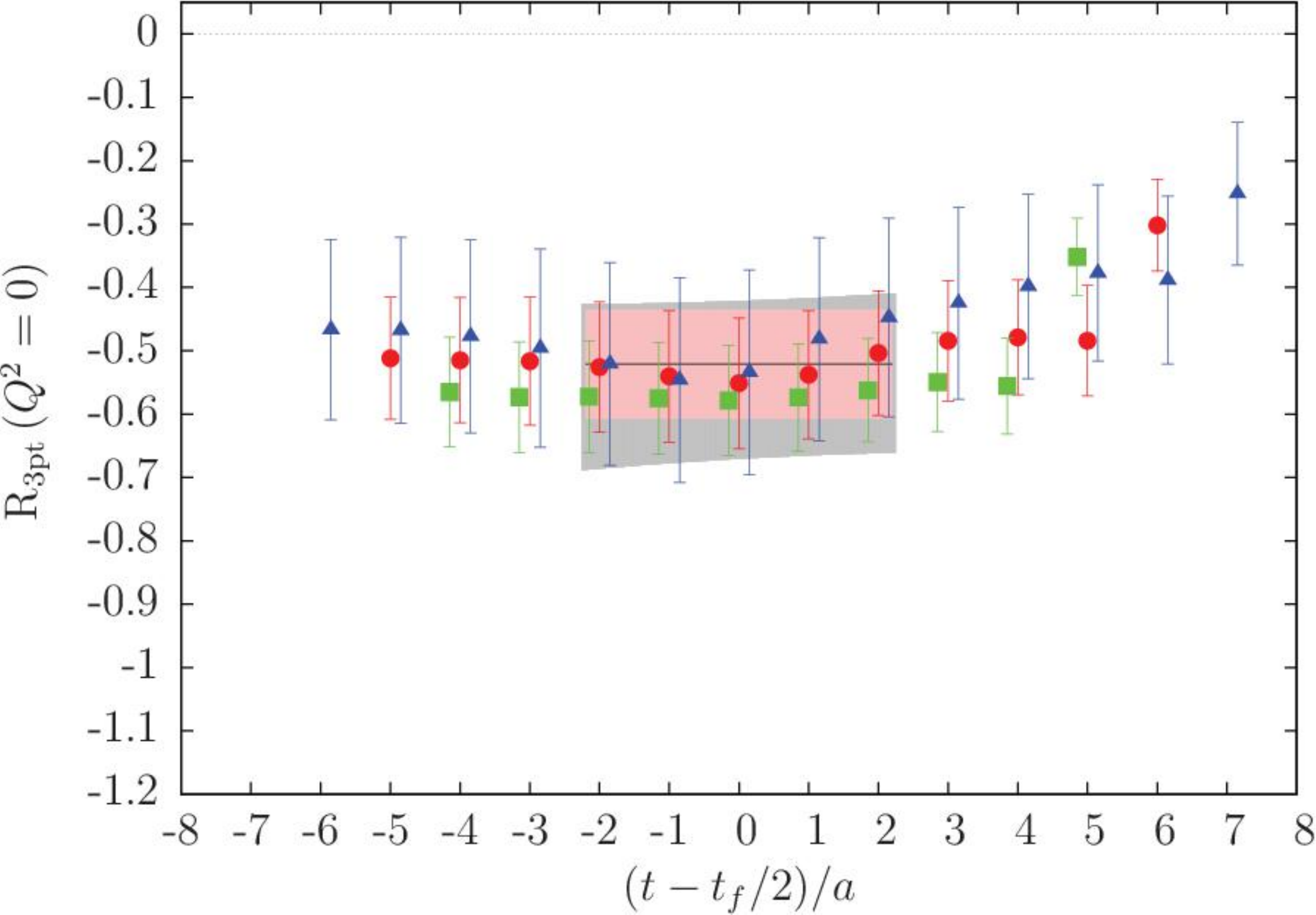} 
\end{minipage}\hfill}
\begin{minipage}{0.425\linewidth}
\includegraphics[width=\linewidth,height=0.78\linewidth]{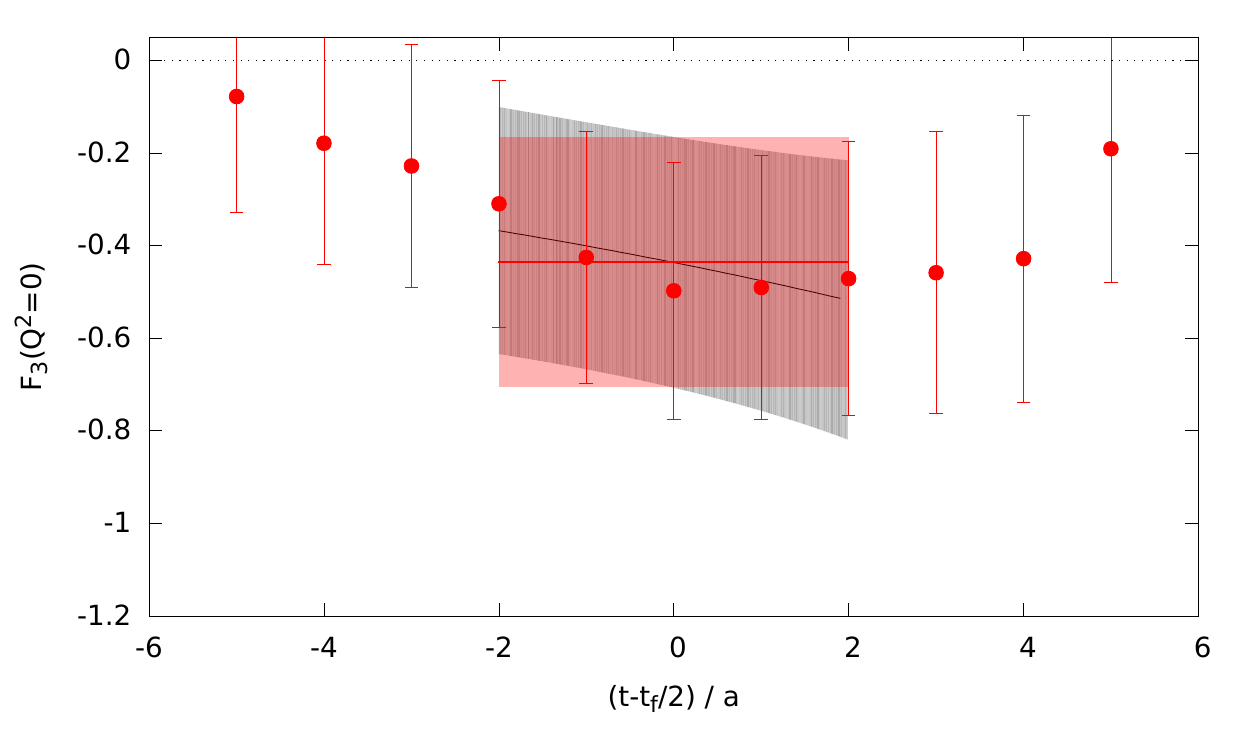} 
\end{minipage}
\caption{Left panel: Computation of $R_{3\text{pt}}(Q^2=0)$ as a function of
the insertion time $\left(t-t_{f}/2\right)/a$ on a $32^3\times 64$ lattice at
$m_\pi=373$~MeV using 4623 statistics at three source-sink separations of
$t_s/a=10,12,14$, denoted by green, red, and blue, respectively.
Right panel: Computation of $F_3(Q^2)$ on a $48^3\times 96$ lattice at
$m_\pi=131$~MeV using 4528 statistics at $t_s/a=10$.}
\label{fig:nedm}
\end{figure}

Despite these successes, there are also challenges ahead.
We demonstrate one such challenge by examining the neutron electric
dipole moment obtained from the CP-odd form factor $F_3(0)$
(see Ref.\ \cite{Alexandrou:2015spa} for technical details).
We show in Fig.~\ref{fig:nedm} simulations for $F_3(Q^2)$ for
three source-sink time separations and two different pion masses and
lattice sizes.
The left panel shows the results for the ratio $R_{3\text{pt}}(Q^2=0)$
(from which $F_3$ can be extracted)
for one ensemble of twisted-mass fermions fitted in the plateau region
with a constant using a pion mass of 373~MeV.
The right panel displays analogous results for $F_3(0)$ using a pion
mass of 131~MeV and the smallest source-sink time separation.
Using the same statistics, and employing the smallest time propagation
$t_s$, the errors turn out to be prohibitively large for the ensemble with a pion
mass close to the physical value.
But a larger time propagation is necessary in order to isolate the nucleon
ground state.
This, however, increases the gauge noise resulting in turn to even larger
errors rendering the determination of $F_3(0)$ computationally very demanding.
Thus, a brute-force approach is not sufficient.
Noise reduction algorithms are needed and a lot of effort has being
devoted towards this goal.

\section{Facilities and Experiments - from Past to Present}
\label{sec:horn}
\footnote{Based on the contribution by T. Horn.}
The study of the internal structure of hadrons is entering a new era. Substantial advances in theory now allow for constructing snapshots of particles that can be compared to experiment. These one-dimensional characteristics can be determined from experiment by extracting the elastic form factors. The resolution of the image is determined by the momentum transfered to the object and new facilities like 12 GeV Jefferson Lab (JLab) enable access to scales thus far not accessible. Form factors thus continue to play an essential role in understanding hadron structure.

Experimental and theoretical knowledge acquired over the past decades, as well as availability of new technology, now enables revolutionary access to a multidimensional representation of the nucleon's inner structure. The pioneering efforts of HERMES and COMPASS, together with the 6 GeV JLab, have demonstrated the feasibility of studying TMDs and to access GPDs through exclusive processes like DVCS (Fig.\ \ref{fig:DVCS}). For example, recent measurements at JLab have demonstrated that high-quality continuous wave (CW) polarized electron beams with a combination of large acceptance and precision detectors are excellent tools for measuring these fundamental distributions. Recent data are also available from RHIC and Drell-Yan studies at Fermilab.

In the near term, the 12 GeV JLab with its extended kinematic range and new experimental hardware has the potential to reveal new aspects of nonperturbative dynamics and the nucleon valence structure. COMPASS can generally provide similar information, but with lower statistical precision and at lower Bjorken $x$ ($x_B$). There will also be data from the Mainz 2 GeV CW microtron facility, which features excellent CW polarized electron beams. However, the lower energy at Mainz will have a very limited kinematic reach as compared to JLab. In the future, the Electron-Ion Collider (EIC) will allow for unprecedented access to the nucleon sea quark and gluon structure. A selection of experiments, projections and results are discussed in the sections below.

\subsection{Measuring elastic form factors}

The electric and magnetic form factors of the nucleon (cf.\ Eqs.\ (2) and (3)) describe the distribution of charge and currents and
are measured using elastic electron scattering (see Fig.\ \ref{fig:scattering}, left panel). At sufficiently small values of the momentum of the virtual-photon probe, $Q^2$, form factors provide a measure of the proton size. The proton size extracted from recent muonic Lamb shift measurements is seven standard deviations smaller than that from electron scattering experiments. This discrepancy known as the ``proton radius puzzle'' illustrates how our limited knowledge of hadron structure also limits high precision tests of QED in atomic systems. The 12 GeV JLab addresses this puzzle with the PRAD electron scattering experiment~\cite{E12-11-106}. PRAD uses a windowless hydrogen gas target and a downstream electromagnetic calorimeter to push to the smallest possible scattering angles, and thus to the smallest possible values of $Q^2$ (down to $Q^2$ = 2 $\times$ 10$^{-4}$ GeV$^2$) to constrain the proton charge radius. The MUSE experiment at the Paul Scherrer Institut (PSI) will address this puzzle with muon scattering.

\begin{figure}[h]                                                      
\centering
\includegraphics[width=7cm,clip]{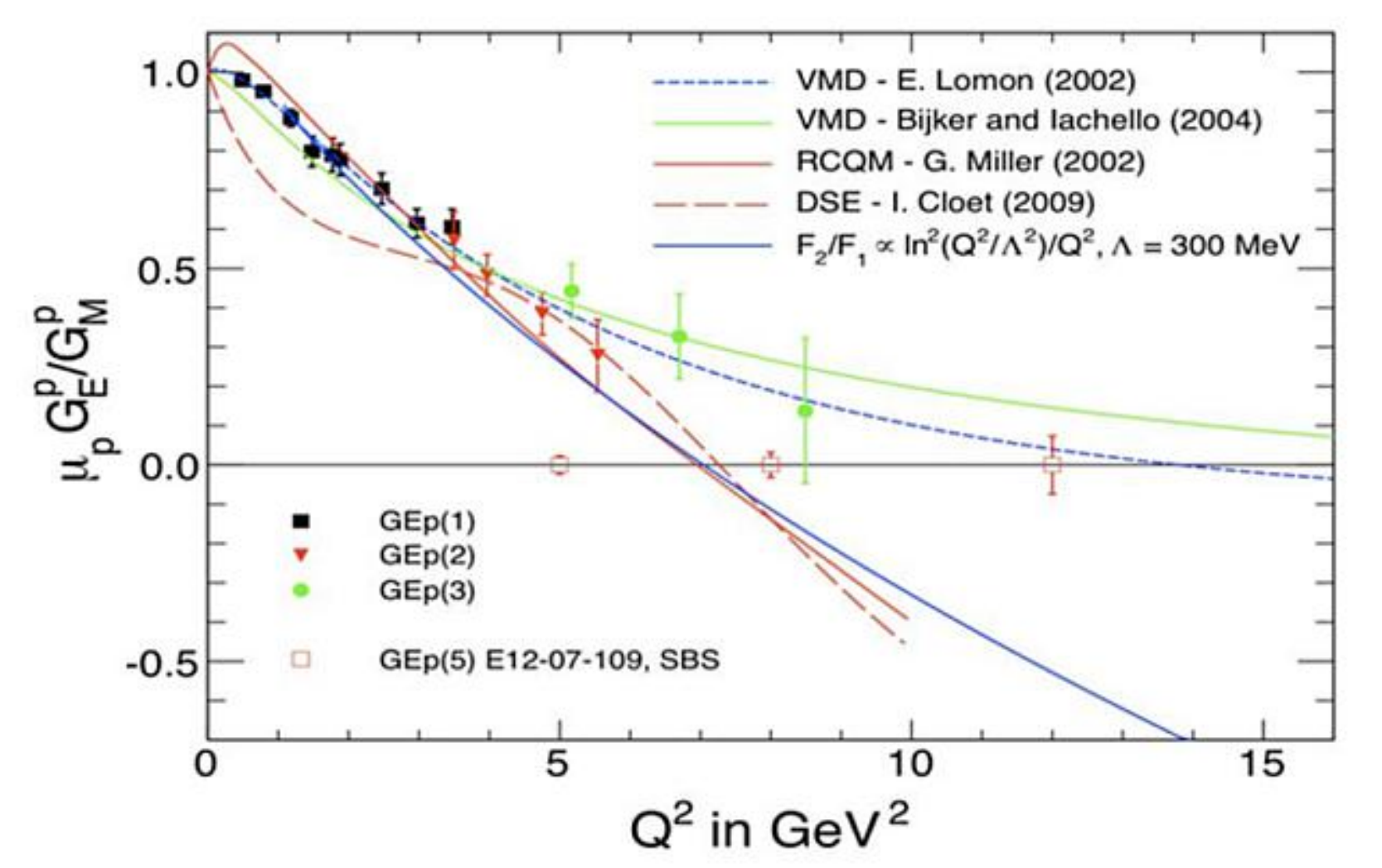} 
\caption{Existing data and projected uncertainties for the ratio of the $\mu_p G_E^p/G_M^p$ Sachs form factors of the nucleon.}
\label{fig-sbs-nucleon-ff}
\end{figure}

Measurements of both proton and neutron form factors over a wide kinematic range allow for flavor separation of charge and magnetization to distance scales deep inside the nucleon. There are a variety of QCD inspired models that can describe the existing form factor data for both the up quark and down quark in the proton at moderate values of $Q^2$. However, these models diverge significantly at larger values of $Q^2$ as illustrated in Fig.~\ref{fig-sbs-nucleon-ff}. The high $Q^2$ region will be probed with high precision by several approved 12 GeV experiments designed to measure the electromagnetic form factors of both the proton and the neutron~\cite{E12-07-108,E12-07-109,E12-07-104,E12-09-019,E12-09-016,E12-11-009}. For the neutron, these new data will nearly triple the range of momentum transfers. The ``Super-Bigbite Spectrometer'' (SBS) in JLab's Hall A with its open geometry and novel GEM detectors will play an important role in the effort to push measurements to the highest possible values of $Q^2 \sim$ 15 GeV$^2$~\cite{Punjabi:2015bba,Wojtsekhowski:2014vua}. These high-precision data will also provide constraints on parameterizations of the GPDs.
\begin{figure}[h]                                                      
\centering
\includegraphics[width=7cm,clip]{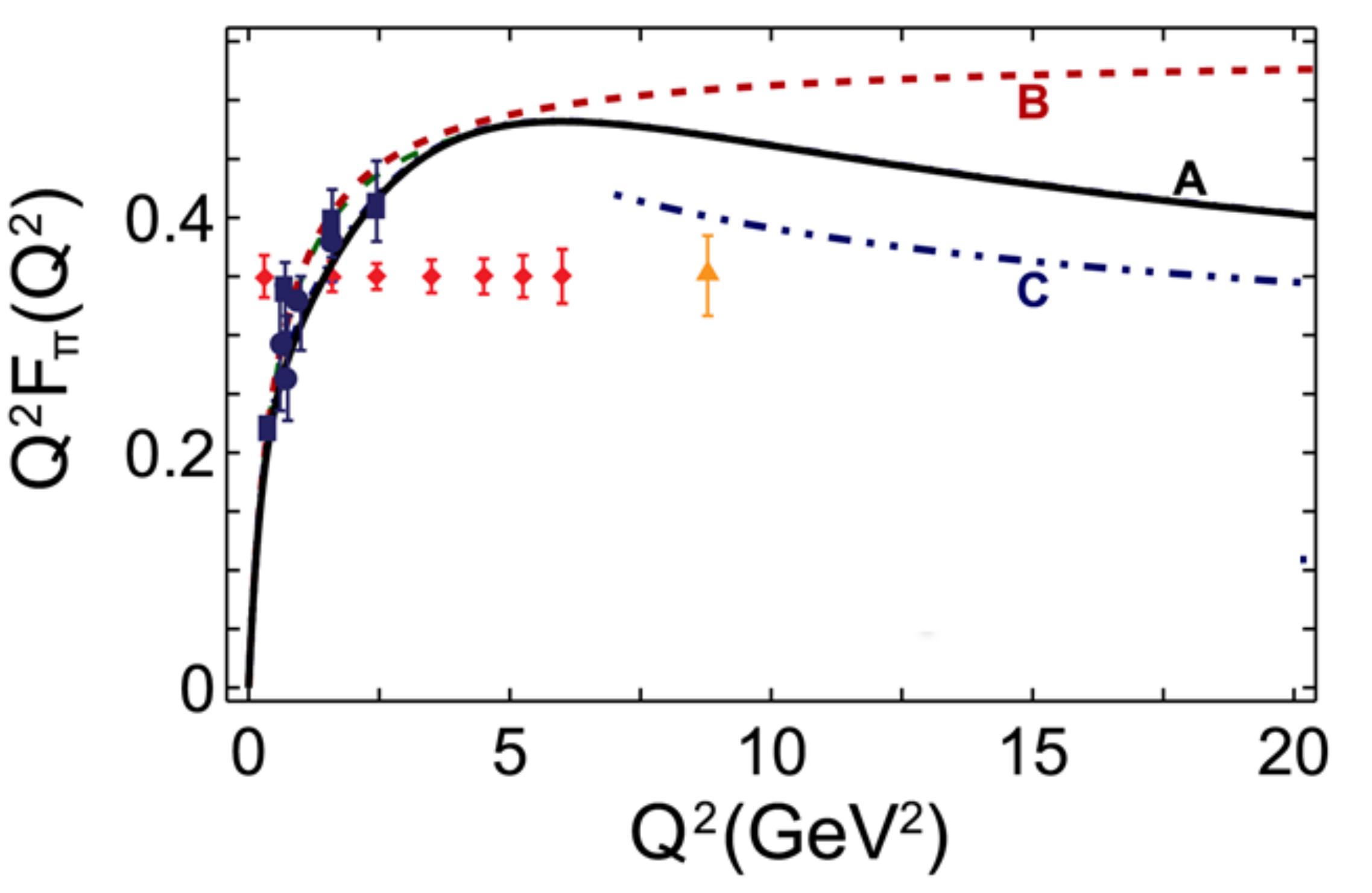} 
\caption{Predictions for $Q^2 F_{\pi}$($Q^2$) vs. existing and projected data. Solid curve (A) shows the prediction of Ref.~\cite{Chang:2013nia}. Remaining curves, from top to bottom: dashed curve (B) monopole form fitted to data in Ref.~\cite{Amendolia:1986wj}, with mass-scale 0.74 GeV; dashed-dotted-dotted curve (C) perturbative QCD prediction using a modern, dilated pion distribution amplitude given by Eq.\ (40) in \cite{Horn:2016rip}. The filled circles and triangle indicate the projected reach and accuracy of forthcoming experiments~\cite{E12-06-101,E12-07-105}.}
\label{fig-shms-hms-pionff}
\end{figure}

The pion is the lightest quark system, with a single valence quark and a single valence antiquark. It is also the particle responsible for the long range character of the strong interaction that binds the atomic nucleus together. A general belief is that the rules governing the strong interaction are left-right, i.e., chirally, symmetric.  If this were true, the pion would have no mass. The chiral symmetry of massless QCD is broken dynamically by quark-gluon interactions and explicitly by inclusion of light quark masses, giving the pion and kaon mass~\cite{Roberts:2016vyn,Horn:2016rip}. The pion and kaon are thus seen as the key to confirm the mechanism that dynamically generates nearly all of the mass of hadrons and central to the effort to understand hadron structure~\cite{Chang:2013nia}.

The 12 GeV JLab 12 features new instrumentation that allows for pushing precision meson form factor measurements to the highest momentum transfers to date. Planned experiments aim for precision measurements of the pion form factor to $Q^2$ = 6 GeV$^2$ and also have the potential to determine the pion form factor up to $Q^2$ 9 GeV$^2$~\cite{E12-06-101,E12-07-105}. Fig.~\ref{fig-shms-hms-pionff} illustrates the kinematic reach that nearly quadruples the range of momentum transfers over which the pion form factor is currently known. These measurements are made possible by the combination of the two moderate acceptance, magnetic spectrometers the Super High Momentum Spectrometer (SHMS) and High Momentum Spectrometer (HMS) pair in JLab's Hall C. The experiments will probe a broad kinematic range over which transition from the large-distance scales with confinement-dominated dynamics at modest $Q^2$ to the short-distance scales with perturbative-dominated dynamics at high $Q^2$ is expected. The data may also shed light on the experimental and theoretical controversy over the large $Q^2$ results for the pion transition form factor~\cite{Aubert:2009mc,Uehara:2012ag} that has refocused attention on the need to understand the distribution of momentum between the valence quark and antiquark.
A classification of theoretical predictions for this transition form factor vs. experimental data, obtained with various pion DAs, can be found in \cite{Bakulev:2012nh}.

\subsection{3D Spatial Mapping - GPDs}

GPDs encode the correlation between the quark/gluon transverse position in the nucleon and its longitudinal momentum, and can be measured directly in exclusive scattering processes at large $Q^2$, in which the nucleon is observed intact in the final state. It is recognized that DVCS and DVMP (illustrated in Fig.\ \ref{fig:DVCS}) are two powerful processes to probe GPDs. Together they offer a path to a full 3D tomography of the nucleon structure.

The key to extracting GPDs from experiment are QCD factorization theorems~\cite{Collins:1996fb,Collins:1998be}, which allow the amplitudes for deep exclusive processes to be expressed in terms of GPDs~\cite{Ji:1996ek,Radyushkin:1996nd,Radyushkin:1996ru,Mueller:1998fv} (discussed in Sec.\ \ref{sec:moutarde}). The value of $Q^2$ at which this formalism is valid experimentally needs to be determined and the contributions of higher twist components to observables need to be quantified.

Deeply Virtual Compton Scattering is the cleanest or golden channel to study GPDs~\cite{Ji:1996ek}. As the DVCS process interferes with the Bethe-Heitler process, one can access the DVCS amplitudes. At leading twist and leading order, one determines Compton Form Factors, which are integrals of GPDs over Bjorken $x$ with a kernel to describe the hard photon-quark interaction. Present analyses assume dominance of several GPDs, validity of twist-2 dominance, and a leading-order formalism. To go beyond this, one has to fully disentangle Compton scattering, Bethe-Heitler contributions, and their interference (after subtracting the known Bethe-Heitler contribution).

The worldwide DVCS experimental program, including experiments at JLab with a 6 GeV electron beam and HERMES with 27 GeV electron and positron beams, has given the first insight into the nucleon GPDs by allowing initial comparisons with models. These experiments have measured large asymmetries, in the 10-20\% range, and suggest an early approach to the hard scattering regime.
\begin{figure}[h]                                                      
\centering
\includegraphics[width=7cm,clip]{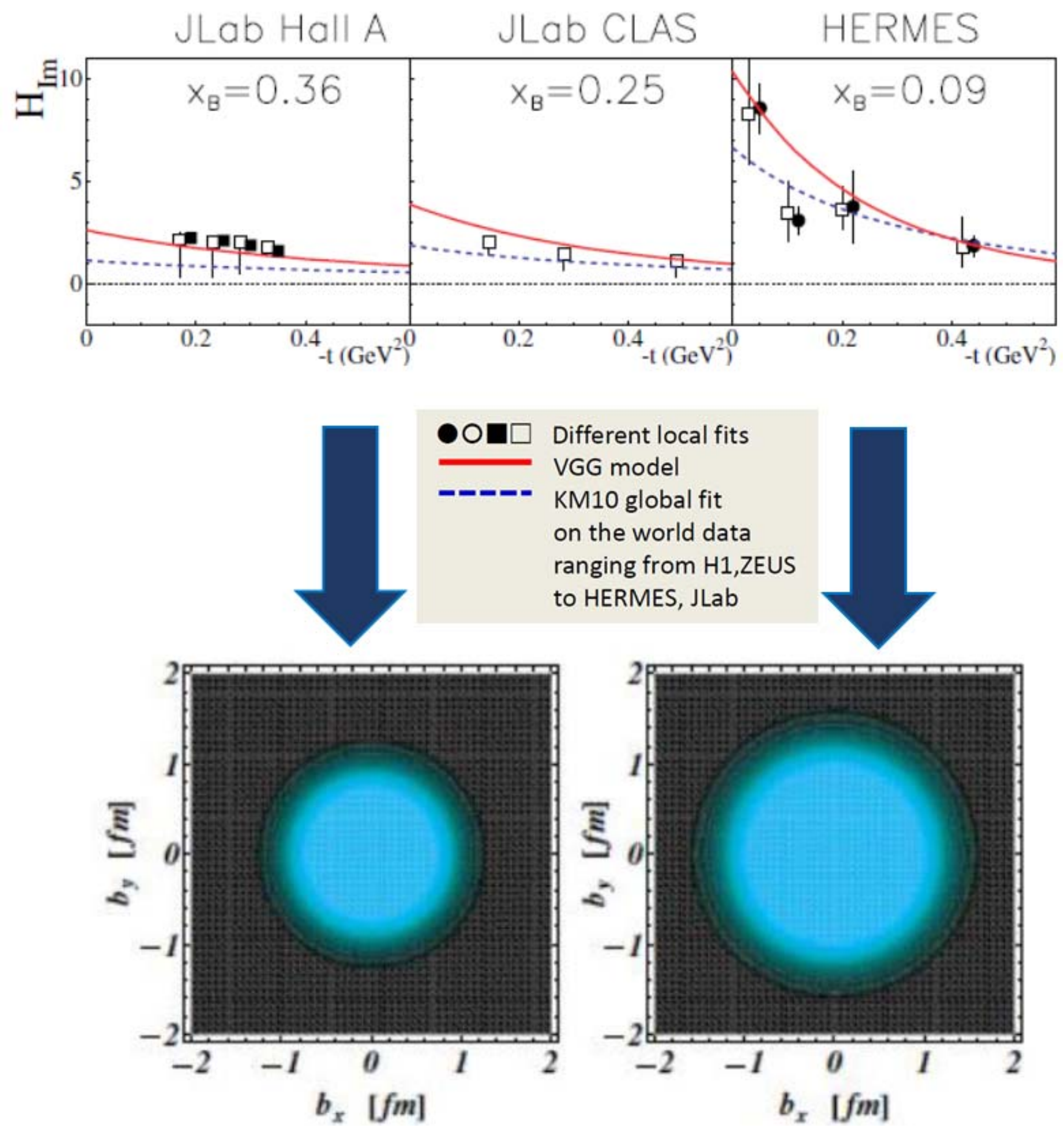} 
\caption{The upper panel shows the determination of the Compton form factor Im{H} in the valence region as a function of $t$ and $x$ using DVCS data collected with CLAS at JLab and HERMES at DESY. The lower panel shows the first 3D views of the nucleon in terms of the spatial charge densities of the proton in a plane ($b_x,b_y$) located at two different values of the quark longitudinal momentum $x$ (taken from \cite{Guidal:2013rya}).}
\label{fig-dvcs-gpd-first-images}
\end{figure}

DVCS cross sections and polarized asymmetries can provide detailed and precise information about GPDs, but are sensitive only to a particular flavor combination, as long as no evolution equations are used. Exclusive meson production provides key additional information allowing the separation of different quark and anti-quark flavors~\cite{Goeke:2001tz}. The theoretical description of these processes is more complicated, and thus measurements that provide information about the reaction mechanism, e.g., tests of hard-soft factorization, are essential. In particular, the emerging transversity GPDs~\cite{Goloskokov:2009ia,Goloskokov:2011rd,Goldstein:2012az} may be accessed if dominance of the transverse cross section at small values of $t<$0.3 GeV$^2$ can be experimentally verified.

To validate the meson factorization theorems and potentially extract flavor separated GPDs from experiment, one has measure the separated longitudinal and transverse (L/T) cross sections and their $t$ and $Q^2$ dependencies. Only L/T separated cross sections can unambiguously show the dominance of longitudinal or transverse photons and allow one to determine possible correlations in $t$ and $Q^2$.The onset of factorization for light mesons may be expected earlier than for heavier ones. Thus, if meson factorization is to be observed it is most probable for pion and kaon.

Accessing GPDs requires a dedicated, long-term experimental effort~\cite{Favart:2015umi}. GPDs are not measured directly, but enter into different combinations and weighted integrals over $x$ as discussed above. To disentangle them requires a diverse array of experiments measuring a variety of observables, including cross sections, beam-spin asymmetries and target-spin asymmetries for both longitudinal and transversely polarized targets. These measurements will be performed for a variety of channels, including both DVCS and DVMP for mesons having different isospins at COMPASS-II at CERN and the 12 GeV JLab.

\begin{figure}[h]                                                      
\centering
\includegraphics[width=7cm,clip]{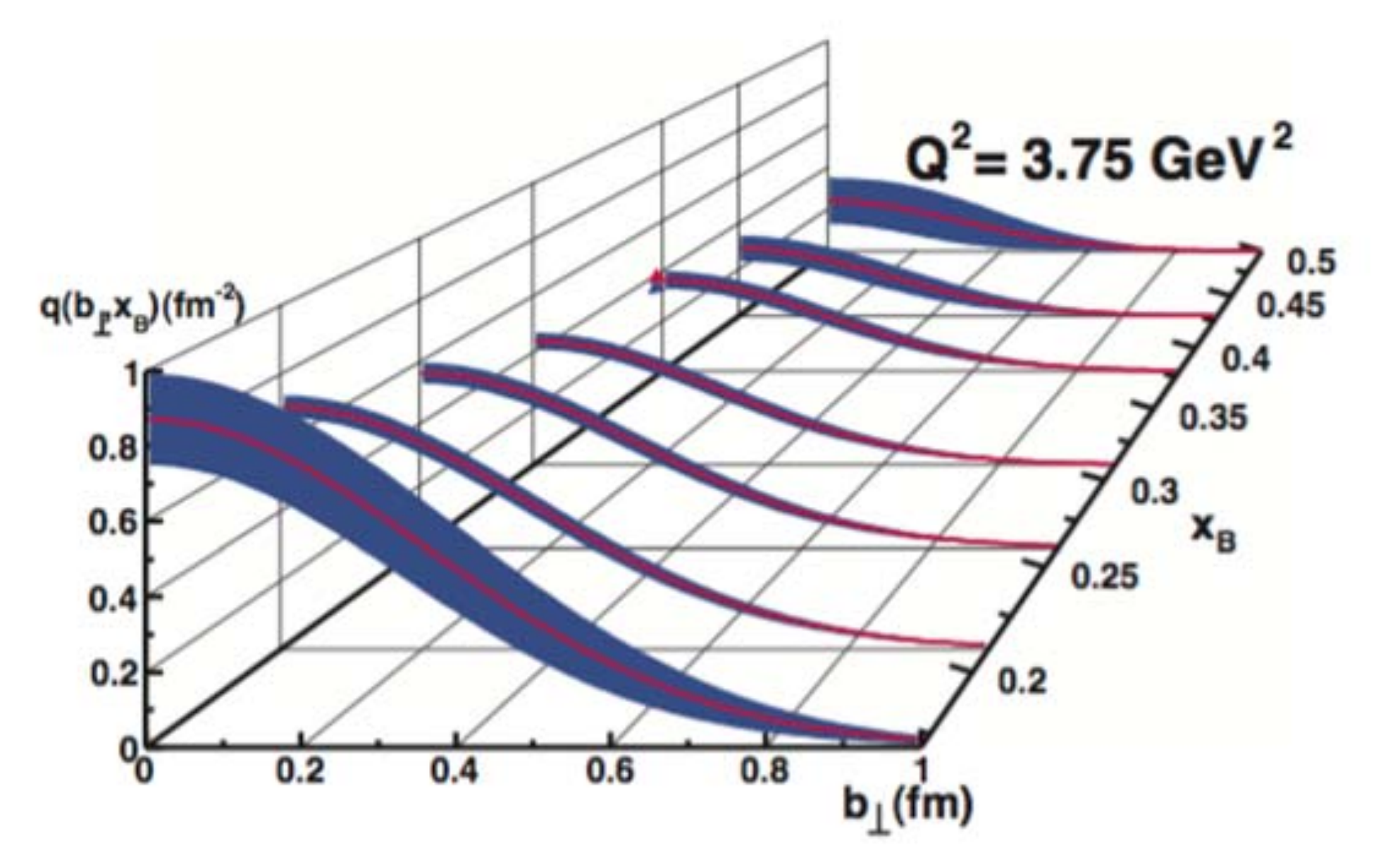} 
\caption{Nucleon transverse profile as function of the impact parameter of a quark relative to the center of the nucleon at fixed $Q^2$ and varying values of $x$. The profile becomes narrower as $x$ increases. The blue band is the projected error centered on the model GPD calculation.}
\label{fig-gpd-transverse}
\end{figure}

The increased energy of the JLab electron beam to 12 GeV offers the kinematic reach where the leading order GPD formalism is anticipated to be applicable. It also provides the highest polarized luminosity for precision measurements of key polarization observables crucial in these studies. A set of approved DVCS experiments planned in Hall B~\cite{E12-06-119,E12-11-003} with CLAS12 and Hall A~\cite{E12-06-114} and Hall C~\cite{E12-13-010} will provide the necessary high-precision data for different channels and reactions over a wide kinematic range. These data will be critical in the extraction of GPDs and parametrizations, while constraints from dispersion-relation techniques and from the lattice calculations of the moments of GPDs, will minimize the model dependence in those parametrizations. Fig.~\ref{fig-gpd-transverse} shows the projected impact of the 12 GeV JLab data on our knowledge of the nucleon transverse spatial profile.

There is an equally ambitious program of experiments involving DVMP, which are able to access GPDs, or combinations thereof, that are inaccessible to DVCS. In Hall B, in parallel with the DVCS measurements of cross sections, structure functions and beam spin asymmetries both for vector mesons and pseudoscalar mesons will be explored over the largest phase space ever probed in the valence regime~\cite{E12-06-108}. Experiments in Hall C will focus on L/T separation for pion electroproduction~\cite{E12-07-105}, and for the first time make precision measurements of $K^+$ cross sections adding strangeness information to the DVMP program~\cite{E12-09-011}. Measuring L/T separated cross sections places strong demands on experimental facilities requiring rigorous control over systematic uncertainties. Hall C at JLab with its precision focusing spectrometers and particle identification detectors is the only facility available for carrying out these measurements.

\subsection{3D Momentum Mapping - TMDs}

One of the impacts of nucleon structure beyond the one-dimensional picture is the introduction of possible orbital motion of partons. Increasingly precise studies of the nucleon spin sum rule by the EMC, E155, HERMES, and STAR collaborations
\cite{Ashman:1987hv,Ashman:1989ig,Anthony:1999py,Anthony:1999rm,Anthony:2000fn,Ackerstaff:1997ws,Airapetian:1998wi,Ackerstaff:1998ja,Adams:2003fx,Adler:2003pb} strongly suggest that the net spin carried by quarks and gluons does not account completely to the net value of the spin of the nucleon, and therefore an orbital angular momentum contribution of partons to the spin of the nucleon must be significant. This in turn implies that transverse momentum of quarks should be non-zero and correlated with the spin of the nucleon itself.

SIDIS is the method of choice to study TMDs~\cite{Mulders:1995dh,Bacchetta:2006tn,Anselmino:2011ay,Cahn:1978se,Cahn:1989yf}.
At leading twist, the dependence of the SIDIS cross section on the azimuthal angle of the electro-produced hadron with respect to the lepton scattering plane and on the nucleon polarization azimuthal angle allows a term-by-term separation of the different azimuthal contributions to the measured unpolarized and polarized cross sections and spin asymmetries. Present analyses rely on the partonic interpretation~\cite{Ji:2004xq} to be valid. The values of $Q^2$, $x$, and $z$ at which this is true have to be experimentally determined. The kinematic dependencies of the basic SIDIS cross section is the only unambiguous way to do that. Basic $\pi^0$ cross sections have experimental advantages to address this.

Among the eight possible TMD functions (see Table \ref{tab:TMDs}), the Sivers function and the Boer-Mulders function have received much theoretical and experimental interest as they are responsible for large observed single-spin asymmetries in SIDIS experiments. Both of these functions are related to the imaginary part of the interference of wave functions having non-zero orbital angular momentum. They describe unpolarized quarks in a transversely polarized nucleon, and transversely polarized quarks in an unpolarized nucleon, respectively. The Sivers transverse momentum distribution function has been recently used~\cite{Bacchetta:2012xf} to infer the GPD $E$ in the collinear limit in order to estimate the angular momentum carried by quarks in the nucleon. In SIDIS the Sivers function is also responsible for ``color lensing'', which describes the overall color attraction between a struck quark on its way to becoming a hadron and the remnant system~\cite{Burkardt:2002ks}. Finally, an important prediction of QCD that needs to be confirmed in experiment is that the Sivers function determined in SIDIS has the opposite sign to that measured in a Drell-Yan experiment. Measurements with both electron beams at Jlab 12 GeV and hadron beams at RHIC, Fermilab, and COMPASS-II at CERN will provide important information.

The multi-dimensional phase space of SIDIS is complex with potentially much unknown physics. With its projected statistical accuracy and the extended kinematic reach the 12 GeV era at Jefferson Lab has the potential to move SIDIS measurements to a new level of sophistication. Each experimental hall brings an essential element to this effort: Hall B with the large-acceptance CLAS12 spectrometer will provide multi-dimensional cross sections, azimuthal distributions and single- and double-spin asymmetries on both polarized and unpolarized neutron and deuteron targets, and on unpolarized nuclear targets. Hall A will provide much needed neutron information through their world-leading polarized 3He target. Finally, Hall C will add precision cross sections and their ratios for both pions and kaons with the SHMS-HMS and Neutral Particle Spectrometer (NPS).
\begin{figure}[h]                                                      
\centering
\includegraphics[width=7cm,clip]{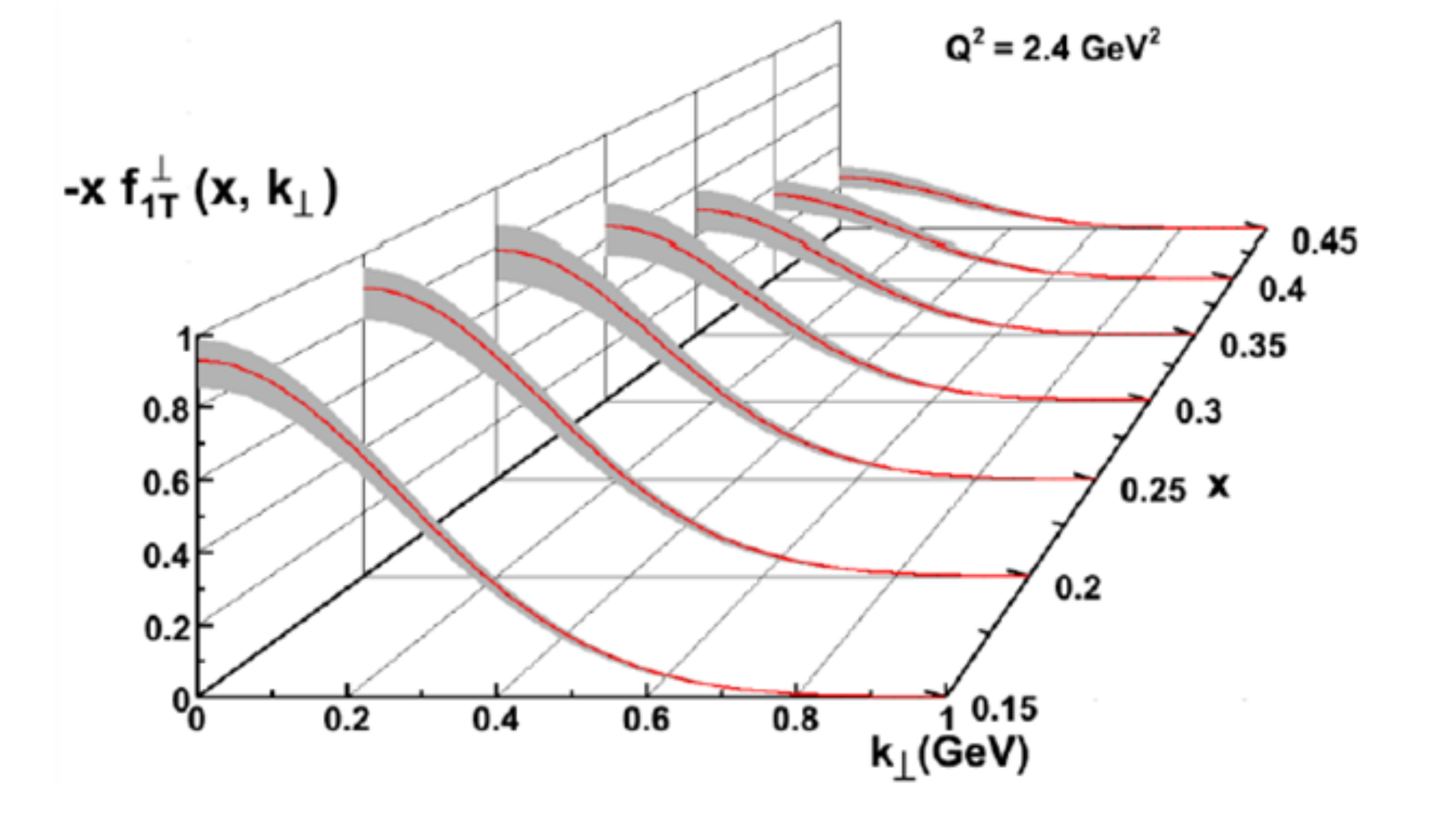} 
\caption{The Sivers function for the up quark as a function of transverse momentum at different values of longitudinal momentum fraction $x$, as projected for JLab 12 GeV with He-3 target. The red line is the model profile of Ref.~\cite{Anselmino:2008sga}. The data to be taken will show the actual shape of the distribution. The gray error bands have been projected around the model profile.}
\label{fig-tmd}
\end{figure}

\begin{figure}[h]                                                      
\centering
\includegraphics[width=7cm,clip]{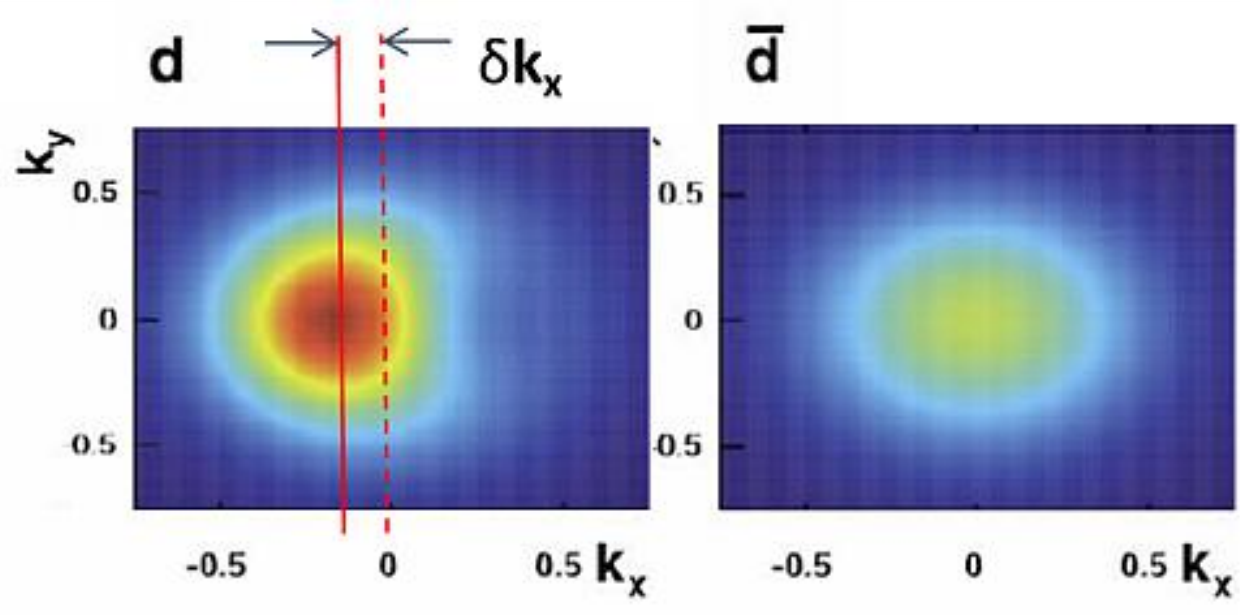} 
\caption{Down quark momentum tomography of the Sivers function at $x$=0.1 for which a non-zero value requires a non-vanishing orbital angular momentum of the quarks. The projected resolution of the JLab 12 GeV experiments shows a clear deformation.}
\label{fig-tmd-d-quark}
\end{figure}

High statistical precision measurements of semi-inclusive pion and kaon production with multi-dimensional binning in the momentum transfer $Q^2$, invariant mass of the unobserved system $W$, the final tagged-hadron energy fraction $z$ and transverse momentum $P_T$ , are essential for performing a model-independent extraction of the TMDs. Such experiments have been proposed and approved for the 12 GeV upgrade. Fig.~\ref{fig-tmd} shows the impact of future data from experiments in Hall B with CLAS12, in Hall A with Super-BigBite~\cite{E12-09-018} and with SoLID~\cite{E12-11-007,E12-11-108} complemented with precision SIDIS experiments in Hall C~\cite{E12-06-104,E12-07-019,E12-13-007} on a precision determination of the Sivers function. These measurements enable a high-resolution tomography of the Sivers function as shown in Fig.~\ref{fig-tmd-d-quark}.

The tensor charge is an important intrinsic property of the nucleon, similar to its axial charge or magnetic
moment, and corresponds to the first moment in $x_B$ of the transversity distribution function $h_1^T$ ($x$). It offers a benchmark test for the most modern lattice QCD calculations. This distribution is accessible in SIDIS, through the well-known Collins effect, by using transversely polarized targets. The tensor charge has been extracted using world data from Hermes, COMPASS and JLab 6 GeV with limited precision. It will be measured in Hall B using CLAS~\cite{E12-11-111} and in Hall A using SoLID~\cite{E12-11-108,E12-10-006} and will be determined with much improved precision in the 12 GeV era.

\section{Facilities and Experiments - the Future} 
\label{sec:EIC}
\footnote{Based on the contribution by T. Horn.}
To understand how the properties and structure of all forms of nuclear matter emerge from the dynamics encoded in QCD, it is essential to precisely image gluons and sea quarks, and to understand the role they and their interactions play in protons, neutrons, and nuclei ~\cite{Boer:2011fh,Accardi:2012qut,Ent:2016lod}. For this, a new accelerator facility is required - the Electron-Ion Collider (EIC). The EIC will exceed the earlier $ep$ collider HERA by providing:
\begin{itemize}
\setlength{\itemindent}{+.5in}
\item{Luminosity a factor of 100-1000 times higher, allowing unprecedented three-dimensional imaging of the gluon and sea quark distributions and to explore correlations among them}
\item{Extensive energy variability to explore the transition in nuclear properties from the region of sea quarks to that of abundant gluons at low $x$, down to 0.001 or 0.0001}
\item{Spin-polarized proton and light ion beams to explore the correlations of gluon and sea quark distributions with the overall nucleon spin, and the contribution of gluons and sea quarks to the nucleon-nucleon interaction}
\item{Heavy-ion beams to reach much higher gluon densities than with proton beams, to study the role and behavior of gluons in nuclei, and to enhance the discovery of collective effects of gluons}
\end{itemize}

The EIC machine designs are described in, e.g. Refs.~\cite{Boer:2011fh,Accardi:2012qut,Ent:2016lod}, and are aimed at achieving
\begin{itemize}
\setlength{\itemindent}{+.5in}
\item{Highly polarized (~70\%) beams of electrons, protons and light nuclei}
\item{Ion beams from deuteron to the heaviest nuclei (uranium or lead)}
\item{Variable $\sqrt{s}$ from $\sim$20 to $\sim$100 GeV, upgradable to $\sim$140 GeV}
\item{High collision luminosity $\sim$10$^{33-34}$ cm$^{-2}$s$^{-1}$}
\item{Possibility to have more than one interaction region}
\end{itemize}

The EIC is designed to provide insight into nucleon structure through multi-dimensional maps of the distributions of partons in space, momentum (including momentum components transverse to the nucleon momentum), spin, and flavor. Measurements enabling the multi-dimensional mapping of the partons, require high luminosity for sufficient statistical precision to be achieved in the multi-dimensional kinematics governing these studies. The 12 GeV JLab~\cite{Dudek:2012vr} and COMPASS at CERN~\cite{Sandacz:2015dqa} will accomplish such studies, predominantly in the valence quark region and somewhat extending into the sea quark region. The EIC will dramatically extend these programs enabling exploration of the role of the gluons and sea quarks in hadron structure and properties.

The EIC will enable measurements of:
\begin{itemize}
\setlength{\itemindent}{+.5in}
\item{the distribution of sea quarks and gluons in momentum and in position space,}
\item{their polarization and their orbital angular momentum, the latter being closely connected with their transverse position and transverse motion since it is a cross product (${\vec L} = {\vec r} \times {\vec p}$),}
\item{correlations between the polarization and distribution of partons in momentum or position space, which may be regarded as the QCD analog of spin-orbit correlations in atomic or nuclear physics.}
\end{itemize}
Such measurements will provide unique information on the \textit{differences} of these distributions when going from small $x$ (a few $\times$ 0.0001) to large $x$ (a few $\times$ 0.1). This will allow one to compare the characteristics of gluons, sea and valence quarks to understand their relation and dynamical interplay, as well as quark flavor dependence. This is of particular interest when comparing light-sea quark flavor distributions, i.e., ${\bar u}$ with ${\bar d}, {\bar s}$ with $({\bar u} + {\bar d})/2$, or $s$ with ${\bar s}$. Significant differences between these distributions are a direct signature of nonperturbative dynamics because perturbative parton radiation is not able to generate them. This re-emphasized special interest in the polarization carried by sea quarks of different flavors, regardless of their contribution to the overall spin of the proton.

\begin{figure}[h]                                                      
\centering
\includegraphics[width=10cm,clip]{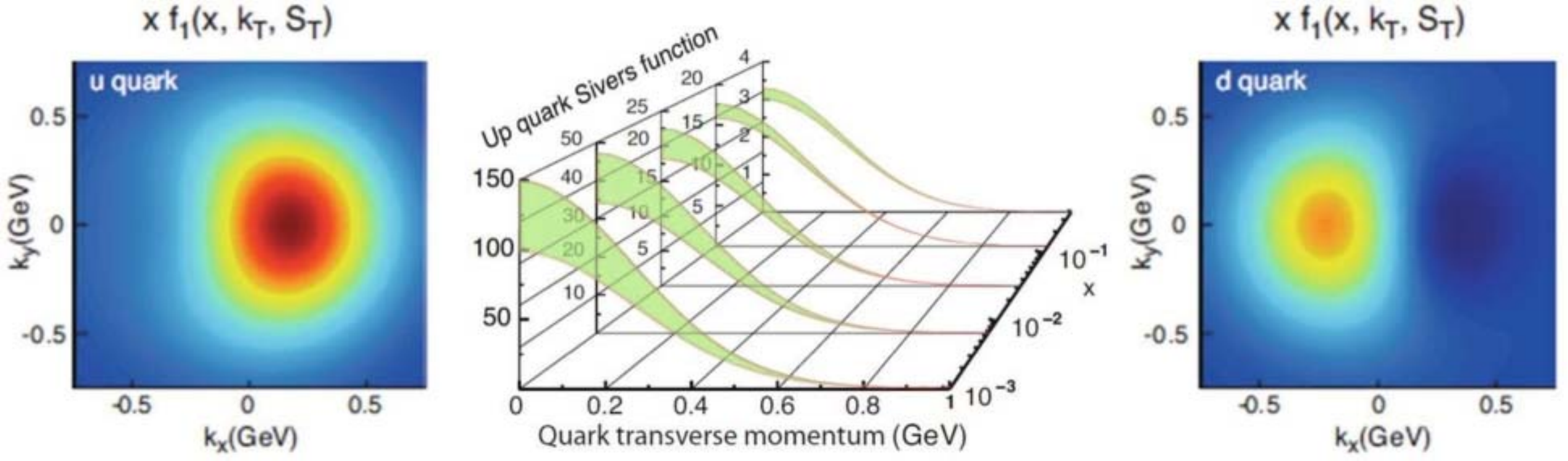} 
\caption{The density in the transverse-momentum plane for unpolarized up quarks (left) and down quarks (right)
with longitudinal momentum fraction $x$ = 0.1 in a transversely polarized proton moving in the $z$-direction,
while polarized in the $y$-direction. The azimuthal asymmetry due to the proton polarization is described by a model
for the Sivers function~\cite{Anselmino:2010bs}.
The color code indicates the probability of finding the up quarks. The
red (blue) shades indicate large negative (positive) values for the Sivers function.
Center: The transverse-momentum profile of the up quark Sivers function at five values of $x$
as accessible to the EIC, and corresponding statistical uncertainties.}
\label{fig-eic-tmd}
\end{figure}

A prominent TMD example, closely connected with the process of dynamical chiral symmetry breaking, is the quark Sivers function $f_1^q(x,\bm{k}_T,S_T)$. It describes how the transverse momentum distribution of unpolarized quarks is correlated with the transverse polarization vector of the nucleon. As a result of this, the quark distribution will be azimuthally asymmetric in the transverse momentum space. Fig.\,\ref{fig-eic-tmd} illustrates the resulting deformations of the up (left) and down (right) quark distributions using a model calculation consistent with current experimental HERMES, COMPASS and JLab data~\cite{Anselmino:2010bs,Anselmino:2011ay}. The center panel of Fig.~\ref{fig-eic-tmd} illustrates the achievable statistical precision of the quark Sivers function from EIC kinematics, ranging from $x \sim$ 0.001 to a few $\times$ 0.1.

There are eight TMDs for gluons. Experimentally, the gluon TMDs and in particular the gluon Sivers function are completely unexplored so far. At an EIC many processes could be used to probe the transverse momentum dependent gluon distributions. One example is electroproduction of a heavy open-charm meson pair ($D{\bar D}$), $\gamma^*N(S_T) \rightarrow D(k_1) + {\bar D}(k_2) + X$, where $N(S_T)$ represents a transversely polarized nucleon, and $D$ and ${\bar D}$ are the two mesons with momenta $k_1$ and $k_2$, respectively. Similar to the Sivers effect in semi-inclusive hadron production, the gluon Sivers function will introduce an azimuthal asymmetry correlating the total transverse momentum $k_T$ of the ($D{\bar D}$) pair with the transverse polarization vector $S_T$ of the nucleon. This will result in a single-spin azimuthal asymmetry. It has been shown that such measurements are feasible at an EIC and in principle have sensitivity to the gluon Sivers function. This would constitute a first measurement of a gluon (TMD) Sivers effect~\cite{Accardi:2012qut}.

With its broad range of collision energies, its high luminosity and nearly hermetic detectors, the EIC could image the proton with unprecedented detail and precision from small to large transverse distances. The accessible parton momentum fractions $x$ extend from a region dominated by sea quarks and gluons to one where valence quarks become important, allowing a connection to the precise images expected from the 12 GeV JLab and COMPASS at CERN. The kinematic access the EIC provides for a range in $x$ transitioning from the region of valence quarks, above $x \sim$ 0.1, through the region where the sea quarks may contribute to nonperturbative nucleon structure, approximately 0.01 $< x \leq$ 0.3, into the region where gluons are abundant, down to about $x$ = 0.0001 or 0.001. To have sufficient resolution, hermeticity and luminosity to measure deep exclusive reactions over this range has been one of the fundamental hypotheses of the EIC design.

The EIC will make possible measurements of deep exclusive processes over an unparalleled large range in $x$ and $Q^2 >$ 10 GeV$^2$. For example, the detection of exclusive $J/\Psi$ meson production would provide unprecedented maps of the gluons transverse spatial distributions within a plane perpendicular to the parent proton motion. Such particular maps encode vital information, inaccessible without EIC, on the amount of proton spin associated with the gluons' orbital motion through the correlation of a longitudinal momentum component $x$ and a transverse spatial position $b_T$.

Measurements of the transverse spatial and momentum distributions of gluons in nuclei could also be performed in a variety of nuclei entering the regime where the onset of collective behavior of gluons, and/or the onset of saturation, is found. This would open a new QCD frontier where the onset towards a universal form of gluon matter with characteristic collective behavior can be observed. In general, the EIC can explore the 3D sea quark and gluon structure of a fast moving nucleus, and verify if such structure differs from that of a free nucleon.
\begin{figure}[h]                                                      
\centering
\includegraphics[width=10cm,clip]{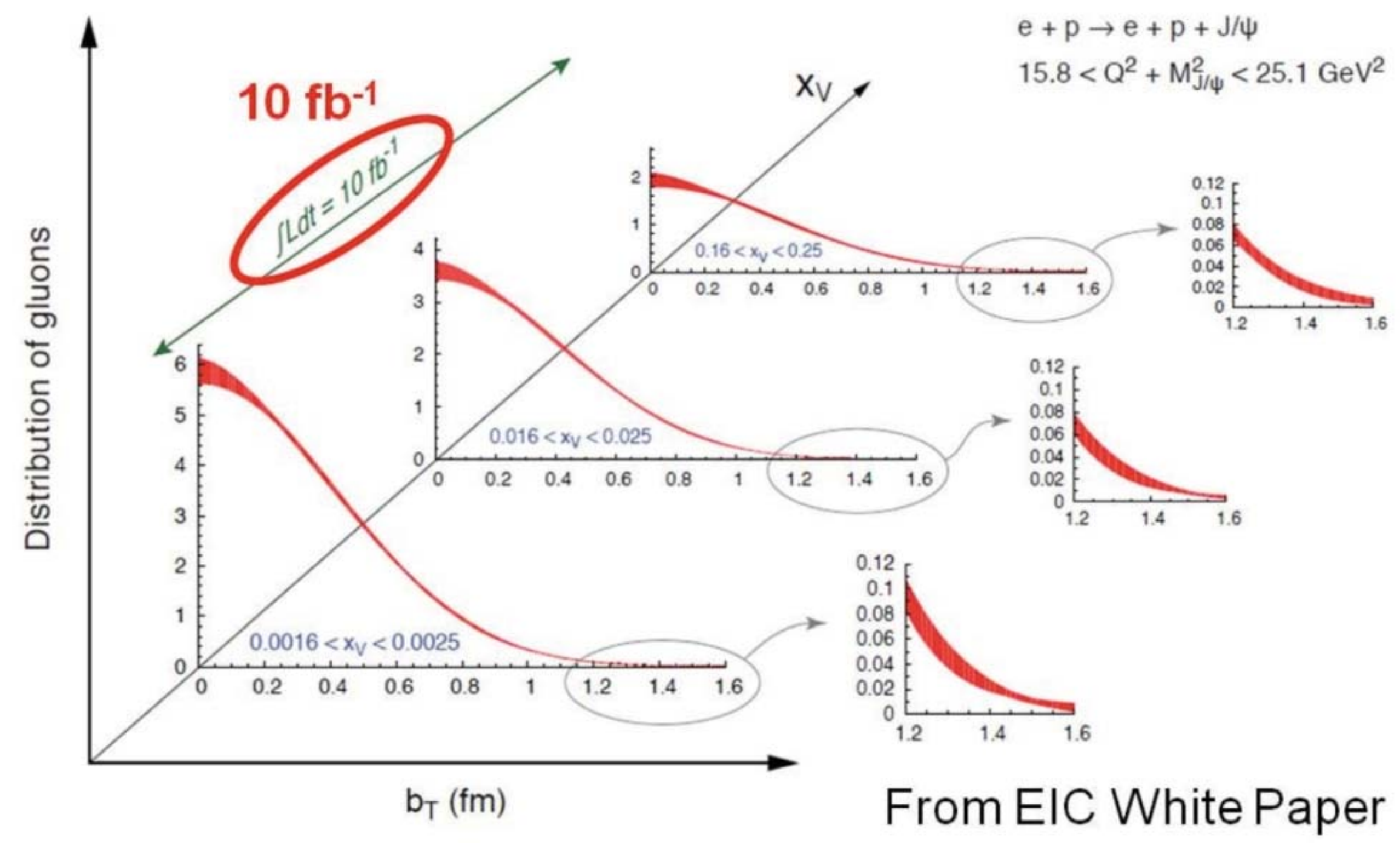} 
\caption{Left panel. Schematic view of a parton with longitudinal momentum fraction $x$ and transverse position
$b_T$ in the proton. Right panel. The projected precision of the transverse spatial distribution of gluons as obtained
from the cross-sections of exclusive $J/\Psi$ production at an EIC~\cite{Accardi:2012qut}.
The distance of the gluon from the center of the proton is $b_T$ in femtometers, and the kinematic
quantity $x_\text{V} = x_\text{B} (1 +M^2_{J/\Psi}/Q^2)$ determines the gluon's momentum fraction.}
\label{fig-eic-gpd}
\end{figure}

With its wide kinematic reach, combined with the capability to probe a variety of nuclei in DIS, semi-inclusive DIS, diffractive and deep exclusive scattering measurements, the EIC allows for exploring the internal 3D sea quark and gluon structure of a fast-moving nucleus. To date it is now known how gluons are distributed in space, for example, if they follow the confinement radius predetermined by quarks, or if they contribute to nuclear structure.  The momentum distributions of gluons in nuclei has been elusive in a similar way. Due to a near-complete lack of experimental constraints, there is not much knowledge about possible modifications of the gluon distributions in nuclei. It is not known if gluons follow the nuclear modifications noted for the quark momentum distributions, known as the nuclear EMC effect. Similarly, in the region of smaller $x$ it remains unclear to what extent the gluons are influenced by shadowing effects noted in ratios of nuclear structure functions.
\begin{figure}[h]                                                      
\centering
\includegraphics[width=10cm,clip]{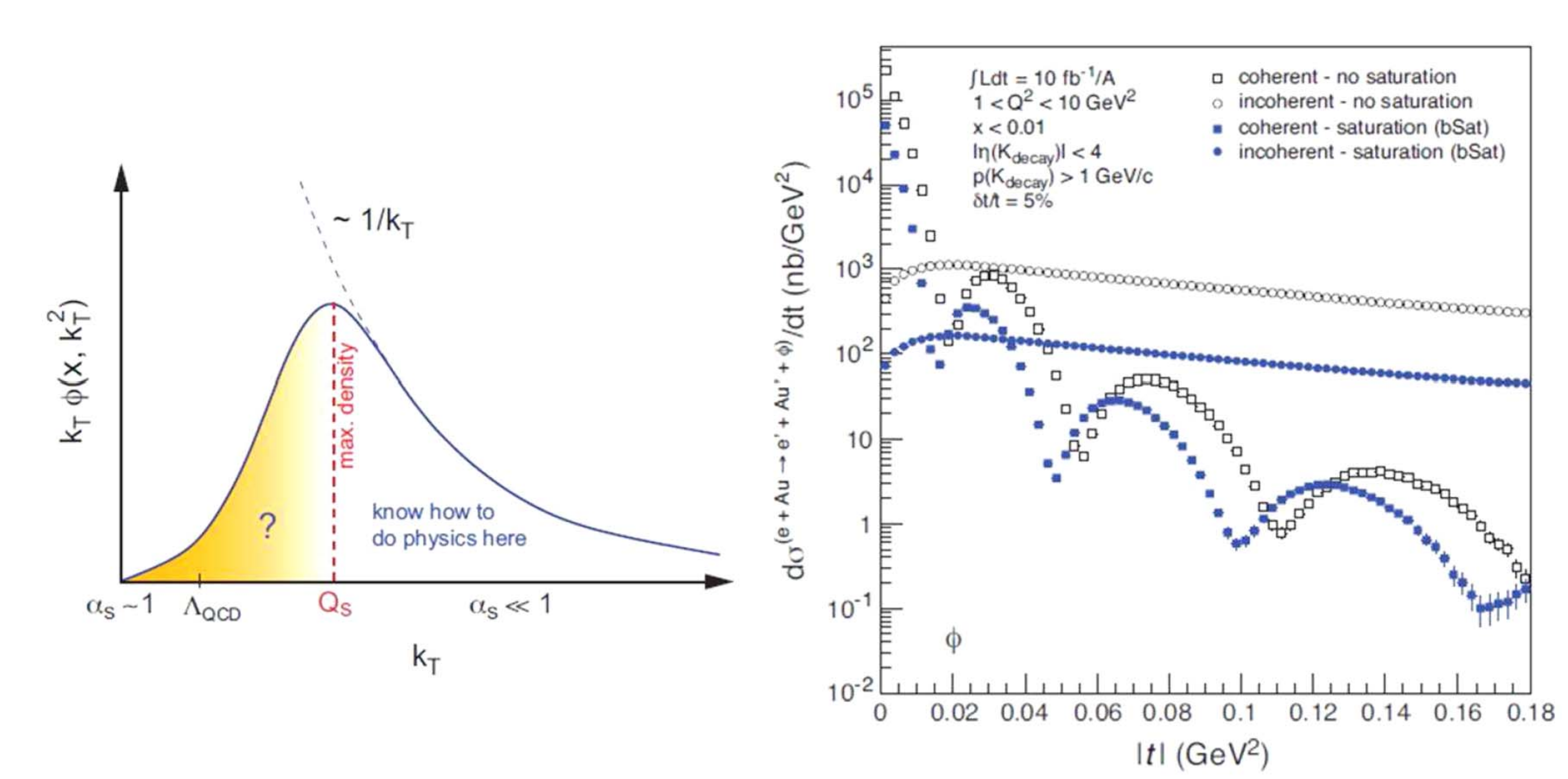} 
\caption{Left panel.  The unintegrated gluon distribution (gluon TMD) of a large nucleus due to
classical gluon fields (solid line). The dashed curve denotes the lowest-order
perturbative result and rises like $\sim 1/k_T$. When the gluon density increases,
this rise must be tamed.
Right panel. Differential cross section $d\sigma/dt$ distributions for exclusive $\phi$ production in
coherent and incoherent events in diffractive e+Au collisions. Predictions from saturation
and non-saturation models are shown in Ref.~\cite{Accardi:2012qut}.}
\label{fig-eic-nuclei}
\end{figure}

With the EIC's excellent hermetic forward detector capabilities for recoil nuclei, extractions of the spatial distributions of gluons in nuclei could be possible from coherent $\phi$ vector meson production in $eA$ scattering. Fig.\,\ref{fig-eic-nuclei} (right) shows an example for diffractive $\phi$ production in an electron-gold scattering process. These measurements could be performed in a variety of nuclei entering the regime where the onset of collective behavior of gluons, and/or the onset of saturation, is found.

\section{Summary and Conclusions}
\label{sec:concl}
Rutherford has laid the foundation for the modern description of the nucleon
and nuclei by inventing the proton and the neutron.
But as it turned out later, also these particles are not elementary bur are
composites of quarks, while the binding dynamics involve also gluons within
the framework of QCD.
For years, one had only a longitudinal perception of the proton, which was
revealed in deep inelastic scattering experiments in terms of longitudinal
momentum PDFs and helicity densities.
Subsequent progress shed light on elastic form factors and the transverse
charge and current densities.
Much more recently, this simple picture was extended significantly to provide
a nucleon image in the transverse plane, thus giving rise to (2+1)D tomography,
parameterized in terms of 3D quark momentum ($\bm{k}_T$) structure (TMDs)
and 3D spatial ($\bm{b}_T$) quark structure (GPDs) functions.
Besides, because quarks and gluons are spin-dependent, one has to understand
how the spin of the proton is distributed among its constituents --- a still
open question.

Several aspects, including TMDs, gluon polarization, etc., have already been
measured at COMPASS-I at CERN (2002-2011), while DVCS, unpolarized SIDIS and TMD
effects, as well as Drell-Yan studies are planned to be carried out in the
COMPASS-II experiment (2012-2017).
The transverse structure of the nucleon in terms of transverse-position
($\bm{b}_T$) dependent GPDs in hard exclusive photon and meson production,
and transverse-momentum ($\bm{k}_T$) dependent TMDs in SIDIS and Drell-Yan
processes will be further explored at COMPASS-II and also at JLab 12 GeV.
The high-energy frontier to study the polarized quark and gluon structure is
provided by the RHIC Spin Program and the planned EIC projects using high-luminosity
polarized beams.
The hope is that all these developments will allow us to obtain a deeper
understanding of the quark-gluon dynamics and their manifestations in the
spatial configuration of the proton structure, both in momentum and
impact parameter space, in the near future.

Bottom line: Our knowledge of the internal structure of hadrons and their properties has evolved continuously. Over the last two decades our
ability to probe the hadron's interior and to develop a clear picture of its internal structure has received a great boost from experiments enabled by modern facilities. A multipronged approach is required, involving constructive feedback between experiment and theory in studies of, e.g. the hadron elastic and transition form factors; generalized and transverse momentum dependent parton distributions. Experimental data from JLab 12 GeV, COMPASS, and the future EIC will be essential in this effort.

\begin{acknowledgement}
All authors of this document would like to thank their respective collaborators.
The work of T. Horn was supported in part by NSF Grant PHY-1306227.
The lattice QCD results reported here (C. Alexandrou) were enabled through a grant
from the Swiss National Supercomputing Centre (CSCS) under project ID s540 as well
as through computational resources  from the John von Neumann-Institute for Computing
on JUROPA and JUQUEEN partly using the PRACE allocation, which included Curie (CEA),
Fermi (CINECA), and SuperMUC (LRZ).
H. Moutarde obtained some of the results and insights described here with collaborators
throughout the world and is grateful to all of them.
This work was supported in part by the Commissariat \`{a} l'Energie Atomique et aux Energies
Alternatives and by the French National Research Agency (ANR) under Grant ANR-12-MONU-0008-01.
I. Scimemi is pleased to acknowledge fruitful discussions with
Miguel G. Echevarria and Alexey Vladimirov.
His work was supported in part by the Spanish MECD Grant FPA2014-53375-C2-2-P.
N. G. Stefanis wants to thank the Alexander von Humboldt-Stiftung for a travel grant
that led to this document.
\end{acknowledgement}


\end{document}